\documentclass[pageno,10pt]{jpaper_arxiv}
\pdfoutput=1

\usepackage[normalem]{ulem}
\usepackage{listings}
\usepackage{protobuf/lang}  
\usepackage{protobuf/style} 
\usepackage{xspace}
\usepackage{tabularx}
\usepackage{booktabs}
\usepackage{amssymb}
\usepackage{amsmath}
\usepackage{multirow}
\usepackage{multicol}
\usepackage{graphicx}
\usepackage[ruled,linesnumbered,noend]{algorithm2e}
\usepackage{authblk}
\usepackage{mathpartir}
\usepackage{balance}
\usepackage{microtype}
\usepackage{protobuf/lang}  
\usepackage{protobuf/style} 

\definecolor{mGreen}{rgb}{0,0.6,0}
\definecolor{mGray}{rgb}{0.5,0.5,0.5}
\definecolor{mPurple}{rgb}{0.58,0,0.82}
\definecolor{backgroundColour}{rgb}{0.95,0.95,0.92}

\lstdefinestyle{CStyle}{
    backgroundcolor=\color{backgroundColour},
    commentstyle=\color{mGreen},
    keywordstyle=\color{magenta},
    numberstyle=\tiny\color{mGray},
    stringstyle=\color{mPurple},
    basicstyle=\footnotesize,
    breakatwhitespace=false,
    breaklines=true,
    captionpos=b,
    keepspaces=true,
    numbers=left,
    numbersep=5pt,
    showspaces=false,
    showstringspaces=false,
    showtabs=false,
    tabsize=2,
    language=C
}

\newtheorem{definition}{Definition}

\newcommand{\revised}[1]{{{#1}}}
\newcommand{\wei}[1]{{{#1}}}

\newcommand{\eva}{EVA\xspace}
\newcommand{\evaend}{EVA\xspace}

\newcommand{\propertyname}[2]{\expandafter\newcommand\csname #1\endcsname[1]{\ensuremath{##1.\mathit{#2}}}}
\propertyname{opcode}{op}
\propertyname{constant}{value}

\newcommand{\evaparam}[2]{\ensuremath{#1.\mathit{parm_{#2}}}}
\newcommand{\evaparams}[1]{\ensuremath{#1.\mathit{parms}}}
\newcommand{\evamodulus}[1]{\ensuremath{#1.\mathit{modulus}}}
\newcommand{\evascale}[1]{\ensuremath{#1.\mathit{scale}}}
\newcommand{\evatype}[1]{\ensuremath{#1.\mathit{type}}}
\newcommand{\evalevel}[1]{\ensuremath{#1.\mathit{level}}}
\newcommand{\evarlevel}[1]{\ensuremath{#1.\mathit{rlevel}}}
\newcommand{\evanpoly}[1]{\ensuremath{#1.\mathit{num\_polynomials}}}

\newcommand{\mathname}[2]{\expandafter\newcommand\csname #1\endcsname{\ensuremath{\mathit{#2}}\xspace}}
\mathname{vecsize}{M}
\mathname{consts}{Consts}
\mathname{insts}{Insts}
\mathname{inputs}{Inputs}
\mathname{outputs}{Outputs}
\mathname{dummy}{id}

\newcommand{\newevaop}[2]{\expandafter\newcommand\csname op#1\endcsname{\ensuremath{\normalfont\textsc{{#2}}}\xspace}}
\newevaop{undefined}{UndefinedOp}
\newevaop{negate}{Negate}
\newevaop{add}{Add}
\newevaop{sub}{Sub}
\newevaop{addsub}{Add/Sub}
\newevaop{multiply}{Multiply}
\newevaop{sum}{Sum}
\newevaop{copy}{Copy}
\newevaop{rotateleft}{RotateLeft}
\newevaop{rotateright}{RotateRight}
\newevaop{relinearize}{Relinearize}
\newevaop{modswitch}{ModSwitch}
\newevaop{rescale}{Rescale}
\newevaop{normalizescale}{NormalizeScale}

\newcommand{\newevatype}[2]{\expandafter\newcommand\csname type#1\endcsname{\ensuremath{\mathbf{#2}}\xspace}}
\newevatype{cipher}{Cipher}
\newevatype{vector}{Vector}
\newevatype{scalar}{Scalar}
\newevatype{integer}{Integer}

\DeclareMathOperator{\plaineval}{\mathcal{E}_{\dummy}}

\graphicspath{{figs/}}

\begin{document}

\title{\evaend: An Encrypted Vector Arithmetic Language and Compiler for Efficient Homomorphic Computation}         


\author[*]{Roshan Dathathri}
\author[$\dag$]{Blagovesta Kostova}
\author[$\ddag$]{Olli Saarikivi}
\author[$\ddag$]{Wei Dai}
\author[$\ddag$]{Kim Laine}
\author[$\ddag$]{Madanlal Musuvathi}
\affil[*]{University of Texas at Austin, USA}
\affil[ ]{\textit {roshan@cs.utexas.edu}}
\affil[$\dag$]{EPFL, Switzerland}
\affil[ ]{\textit {blagovesta.pirelli@epfl.ch}}
\affil[$\ddag$]{Microsoft Research, USA}
\affil[ ]{\textit {\{olsaarik,wei.dai,kilai,madanm\}@microsoft.com}}

\date{}
\maketitle
\thispagestyle{empty}

\begin{abstract}
Fully-Homomorphic Encryption (FHE) offers powerful capabilities by enabling
secure offloading of both storage and computation, and recent innovations in
schemes and implementations have made it all the more attractive. At the same
time, FHE is notoriously hard to use with a very constrained programming model, a
very unusual performance profile, and many cryptographic constraints.
Existing compilers for FHE either target simpler but less efficient FHE schemes
or only support specific domains where they can rely on \wei{expert-provided} high-level runtimes to hide complications.

This paper presents a new FHE language called Encrypted Vector Arithmetic (\evaend), which
includes an optimizing compiler that generates correct and secure FHE programs,
while hiding all the complexities of the target FHE scheme.
Bolstered by our optimizing compiler, programmers can develop efficient general-purpose FHE applications directly in \evaend.
For example, we have developed image processing 
applications using \evaend,
with \wei{a very few} lines of code.

\eva is designed to also work as an intermediate representation 
that can be a target for compiling higher-level domain-specific languages. 
To demonstrate this, we have re-targeted CHET, an existing domain-specific compiler for neural network inference, onto \eva. 
Due to the novel optimizations in \evaend, 
its programs 
are on average $5.3\times$ faster 
than those generated by CHET.
We believe \wei{that} \eva would enable a wider adoption of FHE by making it easier to develop FHE
applications and domain-specific FHE compilers.
\end{abstract}





\section{Introduction}
\label{sec:intro}
Fully-Homomorphic Encryption (FHE) allows arbitrary computations on encrypted data without requiring the decryption key.
Thus, FHE enables interesting privacy-preserving capabilities, such as offloading secure storage and secure computation to untrusted cloud providers. 
Recent advances in FHE theory~\cite{AC:CKKS17,SAC:CHKKS18} along with improved implementations have pushed FHE into the realm of practicality. 
For instance, \wei{with appropriate optimization}, we can perform encrypted fixed-point multiplications 
within a few microseconds,
which matches the speed of 8086 processors that jumpstarted the computing revolution.
Future cryptographic innovations will further reduce the performance gap between encrypted and unencrypted computations. 

Despite the availability of multiple open-source implementations~\cite{sealcrypto,heaangit,palisade,helib}, programming FHE applications remains hard and requires
cryptographic expertise, making it inaccessible to most programmers today. 
Furthermore, different FHE schemes provide subtly different functionalities and require manually setting encryption parameters that control correctness, performance, and security.
We expect the programming complexity to only increase as future FHE schemes become more capable and performant. 
For instance, 
the recently invented CKKS scheme~\cite{AC:CKKS17} supports fixed-point arithmetic operations
by representing real numbers as integers with a fixed scaling factor, 
but requires the programmer to perform rescaling operations 
so that scaling factors and the cryptographic noise 
do not grow exponentially due to multiplications.
Moreover, the so-called RNS-variant of the CKKS scheme~\cite{SAC:CHKKS18} provides efficient implementations that can use machine-sized
integer operations as opposed to multi-precision libraries, but imposes further restrictions on the circuits that can be evaluated on encrypted data.

To improve the developer friendliness of FHE, this paper proposes a new general-purpose language for FHE computation called Encrypted Vector Arithmetic (\evaend). 
\eva is also designed to be an intermediate representation that is a backend for other domain-specific compilers.
At its core, \eva supports arithmetic on fixed-width vectors and scalars. The vector instructions naturally match the encrypted SIMD -- or batching -- capabilities of FHE schemes today. 
\eva includes an optimizing compiler that hides all the complexities of the target FHE scheme, such as encryption parameters and noise.
It ensures that the generated FHE program is correct, performant, and secure. 
\revised{In particular, it eliminates all common runtime exceptions that arise when using FHE libraries. }

\eva implements FHE-specific optimizations, such as 
optimally inserting operations like rescaling and modulus switching.
We have built a compiler incorporating all these optimizations 
to generate efficient programs
that run using the Microsoft SEAL~\cite{sealcrypto} FHE \wei{library which implements}
the RNS-variant of the CKKS scheme.
We have built an \eva executor that 
transparently parallelizes the generated program efficiently, 
allowing programs to scale well.
The executor also automatically reuses the memory used for encrypted messages, 
thereby reducing the memory consumed. 

To demonstrate \evaend's usability, 
we have built a Python frontend for it. 
Using this frontend,
we have implemented several applications in \eva 
with \wei{a very few} lines of code and much \wei{less} complexity 
than in SEAL directly. 
One application computes 
the length of a path in 3-dimensional space, 
which can be used in secure fitness mobile applications. 
We have implemented some statistical machine learning applications. 
We have also implemented two image processing applications, 
Sobel filter detection 
and Harris corner detection.
\revised{We believe Harris corner
detection is one of the most complex programs that \wei{have} been
evaluated using CKKS.}

In addition, we have built 
a domain-specific compiler on top of \eva 
for deep neural network (DNN) inference.
This compiler takes programs written in 
a higher-level language as input 
and generates \eva programs using a library of 
operations on higher-level constructs like 
tensors and images.
In particular, our DNN compiler subsumes the recently proposed domain-specific compiler called CHET~\cite{CHET}. 
Our DNN compiler uses the same tensor kernels as CHET, 
except that it generates EVA programs 
instead of generating SEAL programs.
Nevertheless, the optimizing compiler in \eva is able to outperform CHET in DNN inference 
by $5.3\times$ on average. 

In summary, \eva is a general-purpose language and 
an intermediate representation that improves the 
programmability of FHE applications by 
guaranteeing correctness and security, while outperforming 
current methods. 

The rest of this paper is organized as follows. 
Section~\ref{sec:background} gives background 
on FHE. 
Section~\ref{sec:language} presents the \eva language. 
Section~\ref{sec:compiler} gives an overview of the 
\eva compiler. 
We then describe transformations and analysis in the compiler 
in Sections~\ref{sec:transformations} and~\ref{sec:analysis} 
respectively. 
Section~\ref{sec:apps} briefly describes the domain-specific 
compilers we built on top of \evaend. 
Our evaluation is presented in Section~\ref{sec:eval}. 
Finally, related work and conclusions are presented 
in Sections~\ref{sec:related} and~\ref{sec:conclusions} 
respectively.


\section{Background and Motivation}
\label{sec:background}
In this section, 
we describe FHE (Section~\ref{sec:fhe}) 
and the challenges in using it (Section~\ref{sec:challenges}).
We also describe an implementation of FHE 
(Section~\ref{sec:seal}). 
Finally, we present the threat model assumed in this paper 
(Section~\ref{sec:threat}).

\subsection{Fully-Homomorphic Encryption (FHE)}
\label{sec:fhe}

An FHE scheme includes four stages: key generation, encryption, evaluation, and decryption.
Most of the efficient FHE schemes, for example, BGV~\cite{ITCS:BraGenVai12}, BFV~\cite{FV12}, and CKKS~\cite{AC:CKKS17}, are constructed on the Ring Learning with Errors (RLWE) problem~\cite{EC:LyuPeiReg10}.
At the time of key generation, a polynomial ring of degree $N$ with \wei{integer coefficients} modulo $Q$ must be chosen to represent ciphertexts and public keys according to the security standard~\cite{HomomorphicEncryptionSecurityStandard}.
We call $Q$ the ciphertext modulus.
A message is encoded to a polynomial, and subsequently encrypted with a public key or a secret key to form a ciphertext consisting of two polynomials of degree up to~\wei{$N-1$}.
Encryption also adds to a ciphertext a small random error that is later \wei{removable} in decryption.

\revised{FHE schemes are malleable by design. From the perspective of the user, they offer a way to encrypt integers (or fixed-point numbers in CKKS --- see the next section) such that certain arithmetic operations can be evaluated on the resulting ciphertexts.}
Evaluation primarily includes four operations: addition of ciphertexts, addition of a ciphertext and a plaintext, multiplication of ciphertexts, and multiplication of a ciphertext and a plaintext.
Decrypting (with a secret key) and decoding reveals the message, as if the computation was performed on unencrypted data.

\revised{Many modern FHE schemes also include a SIMD-like feature known as \emph{batching} \wei{which} allows a vector of values to be encrypted as a single ciphertext ($N/2$ values in CKKS). With batching, arithmetic operations happen in an element-wise fashion.
{\it Batching-compatible} schemes can evaluate rotations \wei{which allow} data movement inside \wei{a ciphertext. But} evaluating each rotation step count needs a distinct public key.}

\subsection{Challenges in Using FHE}
\label{sec:challenges}

Programmers using FHE face significant challenges that
must be overcome for correct, efficient, and secure computation. We discuss those
challenges here to motivate our work.

\paragraph{Depth of Computation:}
Computations on ciphertexts increase the initially small error in them linearly on the number of homomorphic additions and exponentially on the multiplicative depth of the evaluation circuit.
When the errors get too large, ciphertexts become corrupted and cannot be decrypted, even with the correct secret key.
\revised{This bound is in turn determined by the size of the encryption parameter $Q$.}
Thus, to support efficient homomorphic evaluation of a circuit, one must optimize the circuit for \revised{low depth.}

\paragraph{Relinearization:}
Each ciphertext consists of $2$ or more polynomials
(freshly encrypted ciphertexts consist of only $2$ polynomials). 
\revised{Multiplication of two ciphertexts with $k$ and $l$ polynomials 
yields a ciphertext with $k+l+1$ polynomials.}
To prevent the number of polynomials from growing
indefinitely, an operation called relinearization is performed to reduce 
it 
back to $2$. 
Relinearization from each distinct number of polynomials to $2$ requires 
a distinct public key.
Relinearization is costly and their optimal placement is an NP-hard problem~\cite{HaoOptRelin}.


\paragraph{CKKS and Approximate Fixed-Point:}
The CKKS~\cite{AC:CKKS17} scheme introduced an additional challenge by only providing
\emph{approximate results} (but much higher performance in return). There are
two main sources of error in CKKS: (i) error from the encoding of values to
polynomials being lossy, and (ii) the noise added in every homomorphic operation
being mixed with the message. To counter this, CKKS adopts a fixed-point
representation \revised{by associating each ciphertext with an unencrypted
scaling factor. Using high enough scaling factors allows the errors to be
hidden.}


CKKS further features an operation called {\it rescaling} that scales down the fixed-point representation of a ciphertext.
Consider a ciphertext $x$ that contains the encoding of $0.25$ multiplied by the scale~$2^{10}$ (a relatively low scale).
$x^2$ encodes $0.0625$ multiplied by the scale~$2^{20}$. Further powers would rapidly overflow modest values of the modulus $Q$, requiring impractically large encryption parameters to be selected.
Rescaling the second power by $2^{10}$ will truncate the fixed-point representation to encode the value at a scale of $2^{10}$.

Rescaling has a secondary effect of also dividing the ciphertext's modulus $Q$
by the same divisor as the ciphertext itself. This means that there is a limited
``budget'' for rescaling built into the initial value of $Q$. The combined
effect for CKKS is that $\log Q$ can grow linearly with the multiplicative depth
of the circuit. It is common to talk about the \emph{level} of a ciphertext as
how much $Q$ is left for rescaling.

A further complication arises from the ciphertext after rescaling being
encrypted under fundamentally different encryption parameters. To apply any
binary homomorphic operations, two ciphertexts must be at the same level, i.e.,
have the same $Q$. Furthermore, addition and subtraction require ciphertexts to
be encoded at the same scale due to the properties of fixed-point arithmetic.
CKKS also supports a modulus switching operation to bring down the level of a
ciphertext without scaling the message. \emph{In our experience, inserting the
appropriate rescaling and modulus switching operations to match levels and
scales is a significantly difficult process even for experts in homomorphic
encryption.}

In the most efficient implementations of CKKS (so called
RNS-variants~\cite{SAC:CHKKS18}), the truncation is actually performed by dividing
the encrypted values by prime factors of~$Q$. Furthermore, there is a fixed
order to these prime factors, which means that from a given level (i.e., how many
prime factors are left in $Q$) there is only one valid divisor available for
rescaling. This complicates selecting points to rescale, as doing so too early
might make the fixed-point representation so small that the approximation errors
destroy the message.

\paragraph{Encryption Parameters:}
In CKKS, all of the concerns about scaling factors, rescaling, and managing levels
are intricately linked with selecting encryption parameters.
Thus, a typical workflow when developing FHE applications
involves a lot of trial-and-error, and repeatedly tweaking the parameters to
achieve both correctness (accuracy) and performance. While some FHE libraries
warn the user if the selected encryption parameters are secure, but not all of
them do, so a developer may need to keep in mind security-related limitations,
which typically means upper-bounding $Q$ for a given~$N$.

\subsection{Microsoft SEAL}
\label{sec:seal}
Microsoft SEAL~\cite{sealcrypto} is a software library that implements the RNS variant of the CKKS scheme.
In SEAL, the modulus $Q$ is a product of several prime factors of bit sizes up to $60$ bits, and
rescaling of ciphertexts is always done by dividing away these prime factors.
The developer must choose these prime factors and order them correctly to achieve the desired
rescaling behavior. SEAL automatically validates encryption parameters for correctness and security.


\subsection{Threat Model}
\label{sec:threat}

We assume a semi-honest threat model, as is typical for homomorphic encryption. This means that
the party performing the computation (i.e., the server) is 
curious about the encrypted data 
but is guaranteed to run the desired operations faithfully. 
This model
matches for example the scenario where the server is trusted, but a malicious party has read access
to the server's internal state and/or communication between the server and the client.

\section{\eva Language}
\label{sec:language}

\begin{table}
\centering
\caption{Types of values.\label{tab:types}}
\begin{tabularx}{\linewidth}{@{}lX@{}}
\toprule
Type & Description \\
\midrule
\typecipher & An encrypted vector of fixed-point values. \\
\typevector & A vector of 64-bit floating point values. \\
\typescalar & A 64-bit floating point value. \\
\typeinteger & A 32-bit signed integer. \\
\bottomrule
\end{tabularx}
\end{table}
        
\begin{table*}
\centering
\caption{Instruction opcodes and their semantics (see Section~\ref{sec:background} for details on semantics of the restricted instructions).\label{tab:opcodes}}
\begin{tabularx}{\linewidth}{@{}llXr@{}}
\toprule
Opcode & Signature & Description & Restrictions \\
\midrule
\opnegate & $\typecipher \to \typecipher$ & Negate each element of the argument. & \\
\opadd & $\typecipher \times (\typevector | \typecipher) \to \typecipher$ & Add arguments element-wise. & \\
\opsub & $\typecipher \times (\typevector | \typecipher) \to \typecipher$ & Subtract right argument from left one element-wise. & \\
\opmultiply & $\typecipher \times (\typevector | \typecipher) \to \typecipher$ & Multiply arguments element-wise (and multiply scales). & \\
\oprotateleft & $\typecipher \times \typeinteger \to \typecipher$ & Rotate elements to the left by given number of indices. & \\
\oprotateright & $\typecipher \times \typeinteger \to \typecipher$ & Rotate elements to the right by given number of indices. & \\
\midrule
\oprelinearize & $\typecipher \to \typecipher$ & Apply relinearization. & Not in input \\
\opmodswitch & $\typecipher \to \typecipher$ & Switch to the next modulus in the modulus chain. & Not in input \\
\oprescale & $\typecipher \times \typescalar \to \typecipher$ & Rescale the ciphertext (and divide scale) with the scalar. & Not in input \\
\bottomrule
\end{tabularx}
\end{table*}
    
The \eva framework uses a single language as its input format, intermediate representation, and executable format. 
\revised{The \eva language abstracts {\it batching-compatible} FHE 
schemes like BFV~\cite{FV12}, BGV~\cite{ITCS:BraGenVai12}, 
and CKKS~\cite{AC:CKKS17,SAC:CHKKS18}, and can be compiled to target 
libraries implementing those schemes.}
Input programs use a subset of the language that omits details specific to FHE, such as when to rescale. 
In this section, we describe the input language and its semantics, while Section~\ref{sec:compiler} presents 
an overview of the compilation to an executable EVA program.

\paragraph{Types and Values:}
Table~\ref{tab:types} lists the types that values in EVA programs may have. The vector types \typecipher and \typevector have a fixed power-of-two size for each input program. The power-of-two requirement comes from the target encryption schemes.

We introduce some notation for talking about types and values in EVA. For \typevector, a literal value with elements $a_i$ is written $[a_1,a_2,\dotsc,a_i]$ or as a comprehension $[a_i\text{ for }i=1\dots i]$. For the $i$th element of \typevector $a$, we write $a_i$. 
For the product type (i.e., tuple) of two EVA types $A$ and $B$, we write $A\times B$, and write tuple literals as $(a,b)$ where $a\in A$ and $b\in B$.

\begin{figure*}
\includegraphics[width=0.98\textwidth]{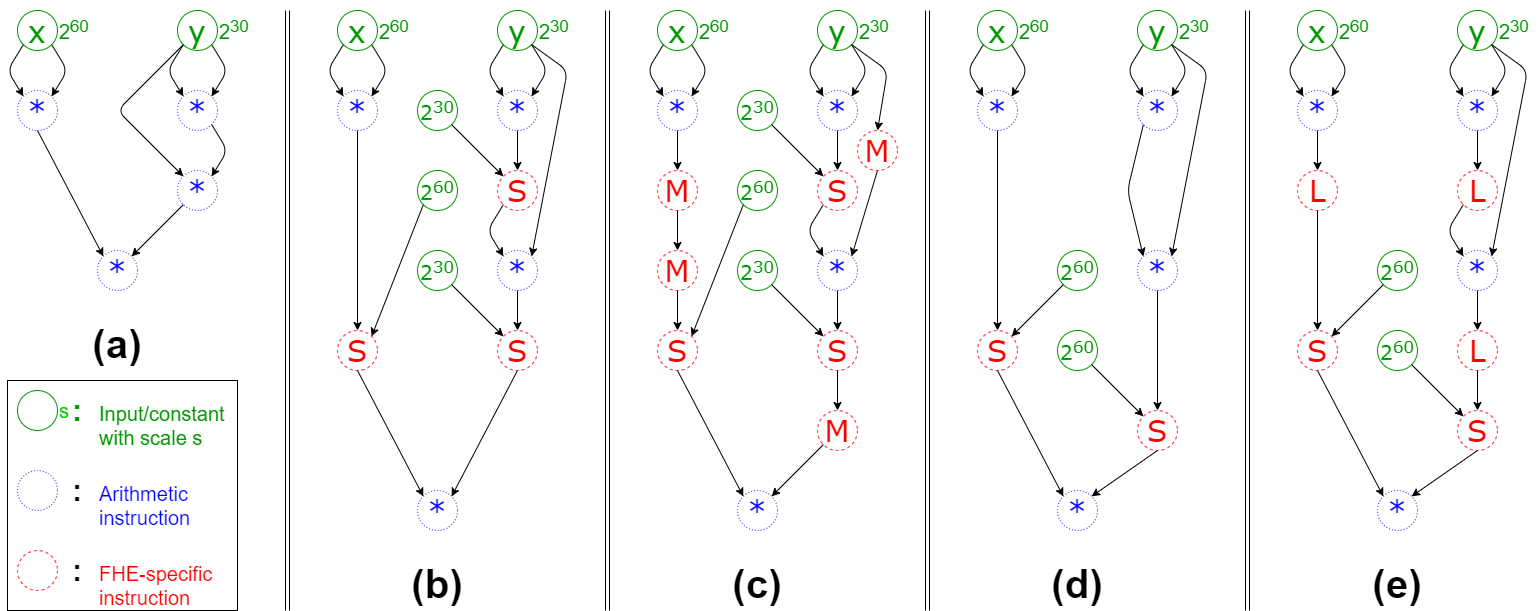}
\caption{\revised{$x^2y^3$ example in \evaend: (a) input; (b) after ALWAYS-RESCALE; (c) after ALWAYS-RESCALE \& MODSWITCH; (d) after WATERLINE-RESCALE; (e) after WATERLINE-RESCALE \& RELINEARIZE (S: \oprescale, M: \opmodswitch, L: \oprelinearize).}}
\label{fig:example1}
\end{figure*}

\paragraph{Instructions:}
Programs in EVA are Directed Acyclic Graphs (DAGs), where each node represents a value available during execution. 
\revised{Example programs are shown in Figures~\ref{fig:example1}(a) and~\ref{fig:example2}(a).}
Nodes with one or more incoming edges are called \emph{instructions}, which compute a new value as a function of its \emph{parameter} nodes, i.e., the parent nodes connected to it. For the $i$th parameter of an instruction $n$, we write $\evaparam{n}{i}$ and the whole list of parameter nodes is $\evaparams{n}$. Each instruction $n$ has an opcode $\opcode{n}$, which specifies the operation to be performed at the node. Note that the incoming edges are ordered, as it corresponds to the list of arguments. Table~\ref{tab:opcodes} lists all the opcodes available in EVA. 
The first group are opcodes that frontends may generate, 
while the second group lists FHE-specific opcodes that are inserted by the compiler. 
\revised{The key to the expressiveness of the input language are 
the \oprotateleft and \oprotateright instructions, which abstract 
rotation (circular shift) in {\it batching-compatible} FHE schemes.}

\paragraph{Inputs:}
A node with no incoming edges is called a \emph{constant} if its value is available at compile time and an \emph{input} if its value is only available at run time. For a constant $n$, we write $\constant{n}$ to denote the value. Inputs may be of any type, while constants can be any type except \typecipher. This difference is due to the fact that the \typecipher type is not fully defined before key generation time, and thus cannot have any values at compile time. The type is accessible as $\evatype{n}$. 

\paragraph{Program:}
A program $P$ is a tuple $(\vecsize,\allowbreak \insts,\allowbreak \consts,\allowbreak \inputs,\allowbreak \outputs)$, where \vecsize is the length of all vector types in $P$; \insts, \consts and \inputs are list of all instruction, constant, and input nodes, respectively; and \outputs identifies a list of instruction nodes as outputs of the program.

\paragraph{Execution Semantics:}
Next, we define execution semantics for EVA. Consider a dummy encryption scheme \dummy that instead of encrypting \typecipher values just stores them as \typevector values. In other words, the encryption and decryption are the identity function. This scheme makes homomorphic computation very easy, as every plaintext operation is its own homomorphic counterpart. Given a map $I: \inputs \to \typevector$, 
\revised{let $\plaineval(n)$ be the function that computes the value for node 
$n$ recursively by using $\constant{n}$ or $I(n)$ if $n$ is a constant or  
input respectively and using $\opcode{n}$ and $\plaineval()$ on $\evaparams{n}$ otherwise}. 
Now for a program $P$, we further define its reference semantic as a function $P_\dummy$, which given a value for each input node maps each output node in $P$ to its resulting value:
\revised{
\begin{align*}
    P_\dummy : \times_{n_i\in\inputs} \evatype{n_i} &\to \times_{n_o\in\outputs} \typevector \\
    P_\dummy(I(n_i^1),\dotsc,I(n_i^{|\inputs|})) &= (\plaineval(n_o^1),\dotsc,\plaineval(n_o^{|\outputs|})) \\
\end{align*}
These execution semantics hold for any encryption scheme, 
except that output is also encrypted (i.e., \typecipher type).
}

\begin{figure*}
\footnotesize
\begin{lstlisting}[language=protobuf3, multicols=2, backgroundcolor=\color{backgroundColour}]
syntax = ``proto3'';

package EVA;

enum OpCode {
  UNDEFINED_OP = 0;
  NEGATE = 1;
  ADD = 2;
  SUB = 3;
  MULTIPLY = 4;
  SUM = 5;
  COPY = 6;
  ROTATE_LEFT = 7;
  ROTATE_RIGHT = 8;
  RELINEARIZE = 9;
  MOD_SWITCH = 10;
  RESCALE = 11;
  NORMALIZE_SCALE = 12;
}

enum ObjectType {
  UNDEFINED_TYPE = 0;
  SCALAR_CONST = 1;
  SCALAR_PLAIN = 2;
  SCALAR_CIPHER = 3;
  VECTOR_CONST = 4;
  VECTOR_PLAIN = 5;
  VECTOR_CIPHER = 6;
}

message Object {
  uint64 id = 1;
}

message Instruction {
  Object output = 1;
  OpCode op_code = 2;
  repeated Object args = 3;
}

message Vector {
  repeated double elements = 1;
}

message Input {
  Object obj = 1;
  ObjectType type = 2;
  double scale = 3;
}

message Constant {
  Object obj = 1;
  ObjectType type = 2;
  double scale = 3;
  Vector vec = 4;
}

message Output {
  Object obj = 1;
  double scale = 2;
}

message Program {
  uint64 vec_size = 1;
  repeated Constant constants = 2;
  repeated Input inputs = 3;
  repeated Output outputs = 4;
  repeated Instruction insts = 5;
}
\end{lstlisting}
\caption{The EVA language definition using Protocol Buffers.}
\label{fig:language}
\end{figure*}

\paragraph{Discussion on Rotation and Vector Sizes:} 
The EVA language restricts the size for any \typecipher or \typevector input 
to be a power-of-2 so as to support execution semantics of 
\oprotateleft and \oprotateright instructions. 

The target encryption schemes use the same vector size $s_e$ ($= N/2$) 
for all \typecipher values during execution. 
However, the vector size $s_i$ for an input could be different from $s_e$. 
Nevertheless, the target encryption scheme and the EVA language enforce 
the vector sizes $s_e$ and $s_i$, respectively, to be powers-of-2.
\eva chooses encryption parameters for the target encryption scheme 
such that for all inputs $i$, $s_i \leq s_e$ 
(because $N$ can be increased trivially without hurting correctness or security 
--- note that increasing $N$ will hurt performance). 

For a \typecipher or \typevector input $i$ such that $s_i < s_e$, 
\eva constructs a vector (before encryption for \typecipher) $i'$ 
with $s_e/s_i$ copies of the vector $i$ contiguously 
such that $s_i' = s_e$, 
and replaces $i$ with $i'$ in the program. 
For example, if an input $a = [a_1,a_2]$ and $s_e = 4$, 
then \eva constructs $a' = [a_1, a_2, a_1, a_2]$ 
and replaces $a$ with $a'$. 
\oprotateright or \oprotateright instruction on the original vector $i$ and 
the constructed larger vector $i'$ 
have the same result on their first $s_i$ elements. 
This is feasible because $s_i$ divides $s_e$. 
\eva thus ensures that the execution semantics 
of the \eva program holds for any encryption scheme.

\paragraph{Implementation:}
As shown in Figure~\ref{fig:language},
the EVA language has a serialized format defined using Protocol Buffers~\cite{protocolbuffer}, a language and platform neutral data serialization format.
Additionally, the EVA language has an in-memory graph representation that is designed for efficient analysis and transformation, which is discussed in Section~\ref{sec:compiler}.

\section{Overview of \eva Compiler}
\label{sec:compiler}


In this section, 
we describe how to use the \eva compiler (Section~\ref{sec:using-eva}).
We then describe the constraints on the code 
generated by \eva (Section~\ref{sec:constraints}). 
Finally, we give an overview of the 
execution flow of the \eva compiler (Section~\ref{sec:eva-flow}).

\subsection{Using the Compiler}
\label{sec:using-eva}

\revised{In this paper, we 
present the \eva compiler for the 
RNS variant of the CKKS scheme~\cite{SAC:CHKKS18} 
and its implementation in the SEAL library~\cite{sealcrypto}. 
Targeting 
\eva for other FHE libraries~\cite{heaangit,helib,palisade} 
implementing CKKS~\cite{AC:CKKS17,SAC:CHKKS18} 
would be straightforward.
The \eva compiler can also be adapted to support 
other {\it batching-compatible} FHE schemes like BFV~\cite{FV12} 
and BGV~\cite{ITCS:BraGenVai12}.}

The \eva compiler takes a program in the \eva language as input. 
Along with the program, it needs the 
fixed-point scales or precisions for each input in the program 
and the desired fixed-point scales or precisions 
for each output in the program. 
The compiler then generates a program in the \eva language as output. 
In addition, it generates 
a vector of bit sizes
that must be used to generate the encryption parameters 
as well as 
a set of rotation steps that must be used to generate the rotation keys. 
The encryption parameters and the rotations keys thus generated 
are required to execute the generated \eva program. 

    
While the input and the output programs are in the \eva language, 
the set of instructions allowed in the input and the output are distinct, 
as listed in Table~\ref{tab:opcodes}. 
The \oprelinearize, \oprescale, and \opmodswitch instructions 
require understanding the intricate details of the FHE scheme. 
Hence, they are omitted from the input program. 
Note that we can make these instructions optional in the input 
and the compiler can handle it if they are present, 
but for the sake of exposition, 
we assume that the input does not have these instructions.

The input scales and the desired output scales 
affect the encryption parameters, and consequently, 
the performance and accuracy of the generated program. 
Choosing the right values for these is a trade-off 
between performance and accuracy 
(while providing the same security). 
Larger values lead to larger encryption parameters and 
more accurate but slower generated program, whereas 
smaller values lead to smaller encryption parameters and 
less accurate but faster generated program. 
Profiling techniques like those used in prior work~\cite{CHET} 
can be used to select the appropriate values.

\subsection{Motivation and Constraints}
\label{sec:constraints}

There is a one-to-one mapping between instructions in 
the \eva language (Table~\ref{tab:opcodes}) 
and instructions in the RNS-CKKS scheme.
However, the input program cannot be directly executed. 
Firstly, encryption parameters are required 
to ensure that the program would be accurate. 
\eva can simply determine the bit sizes that 
is required to generate the parameters. 
However, this is insufficient to execute the program 
correctly because some instructions in the RNS-CKKS 
scheme have restrictions on their inputs. 
If these restrictions are not met, 
the instructions would just throw an exception at runtime. 
    
Each ciphertext in the RNS-CKKS scheme has a coefficient 
modulus \revised{$q$ ($Q = \prod_{i=1}^{r} q_{i}$)}\footnote{In SEAL, 
if the coefficient modulus $q$ is $\{q_1, q_2, ..., q_r\}$, then 
$q_i$ is a prime close to a power-of-2. 
\eva compiler 
(and the rest of this paper) 
assumes 
$q_i$ is the corresponding power-of-2 instead. 
To resolve this discrepancy, 
when a \oprescale instruction divides the scale by the prime, 
the scale is adjusted (by the \eva executor) 
as if it was divided by the power-of-2 instead.} 
and a
fixed-point $scale$ associated with it. 
\revised{All freshly encrypted ciphertexts have the same $q$ 
but they may have different $scale$.}
The following constraints apply for the
binary instructions involving two ciphertexts:
\begin{equation}\label{eq:constraint-mod}
\begin{array}{c}
\evamodulus{\evaparam{n}{1}} = \evamodulus{\evaparam{n}{2}}  \\
\text{ if } \opcode{n} \in \{\opadd,\opsub,\opmultiply\}
\end{array}
\end{equation}
\begin{equation}\label{eq:constraint-scale}
\begin{array}{c}
\evascale{\evaparam{n}{1}} = \evascale{\evaparam{n}{2}} \\
\text{ if } \opcode{n} \in \{\opadd,\opsub\}
\end{array}
\end{equation}
In the rest of this paper, 
whenever we mention \opadd regarding constraints, 
it includes both \opadd and \opsub.

\revised{Consider the example to compute $x^2y^3$ 
for ciphertexts $x$ and $y$ in Figure~\ref{fig:example1} 
(viewed as a dataflow graph).
Constraint~\ref{eq:constraint-scale} is trivially satisfied 
because they are no \opadd instructions. 
Only \oprescale and \opmodswitch instructions modify $q$. 
Constraint~\ref{eq:constraint-mod} is also trivially satisfied 
due to the absence of these instructions.}
Nonetheless, without the use of \oprescale instructions, 
the scales and the {\it noise} 
of the ciphertexts would grow exponentially 
with the multiplicative depth of the program 
(i.e., maximum number of \opmultiply nodes in any path) 
and consequently, 
the $\log_2$ of the coefficient modulus product $Q$ required for the input 
would grow exponentially. 
Instead, using \oprescale instructions ensures that 
$\log_2 Q$ would only grow linearly 
with the multiplicative depth of the program. 

\revised{In Figure~\ref{fig:example1}, 
the output has a scale of $2^{60}*2^{60}*2^{30}*2^{30}*2^{30}$ 
(with $\evascale{x} = 2^{60}$ and $\evascale{y} = 2^{30}$). 
This would require $Q$ to be at least $2^{210}*s_o$, 
where $s_o$ is the user-provided desired output scale. 
To try to reduce this, one can insert \oprescale after every \opmultiply, 
as shown in Figure~\ref{fig:example1}(b). 
However, this yields an invalid program because 
it violates Constraint~\ref{eq:constraint-mod} for the last (bottom) \opmultiply
(and there is no way to choose the same $q$ for both $x$ and $y$). 
To satisfy this constraint, 
\opmodswitch instructions can be inserted, 
as shown in Figure~\ref{fig:example1}(c). 
Both \oprescale and \opmodswitch drop the first element in their input $q$ 
(or {\it consume the modulus}), 
whereas \oprescale also divides the scale by the given scalar 
(which is required to match the first element in $q$). 
The output now has a scale of $2^{60}*2^{30}$. 
This would require choosing $q = \{2^{30},2^{30},2^{60},2^{60},2^{30},s_o\}$. 
Thus, although Figure~\ref{fig:example1}(c) executes more instructions 
than Figure~\ref{fig:example1}(a), 
it requires the same $Q$.
A better way to insert \oprescale instructions is shown in 
Figure~\ref{fig:example1}(d). 
This satisfies Constraint~\ref{eq:constraint-mod} without \opmodswitch instructions.
The output now has a scale of $2^{60}*2^{30}$. 
We can choose $q = \{2^{60},2^{60},2^{30},s_o\}$, 
so $Q = 2^{150}*s_o$. 
Hence, this program is more efficient than the input program.

\begin{figure}
\includegraphics[width=0.47\textwidth]{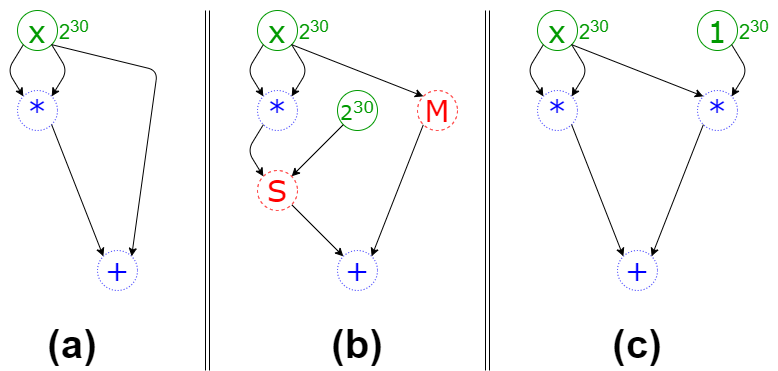}
\caption{\revised{$x^2+x$ example in \evaend: (a) input; (b) after ALWAYS-RESCALE \& MODSWITCH; (c) after MATCH-SCALE.}}
\label{fig:example2}
\end{figure}
    
If the computation was modified to $x^2 + y^3$ in Figure~\ref{fig:example1}(d), 
then the last (bottom) \opmultiply would be replaced by \opadd 
and the program would violate Constraint~\ref{eq:constraint-scale} 
as \opadd would have operands with scales $2^{60}$ and $2^{30}$. 
Consider a similar but simpler example of $x^2 + x$ in Figure~\ref{fig:example2}(a).
One way to satisfy Constraints~\ref{eq:constraint-mod} and~\ref{eq:constraint-scale} 
is by adding \oprescale and \opmodswitch,
as shown in Figure~\ref{fig:example2}(b), 
which would require $q = \{2^{30},2^{30},s_o\}$. 
Another way is to introduce \opmultiply 
of $x$ and a constant 1 with $2^{30}$ scale 
to match the scale of \opadd operands, 
as shown in Figure~\ref{fig:example2}(c), 
which would require $q = \{2^{60},s_o\}$. 
Although the product $Q = 2^{60}*s_o$ is same in both cases, 
the modulus length $r$ is different. 
Hence, the program in Figure~\ref{fig:example2}(c) is more efficient 
due to a smaller $r$.
}

\revised{\opmultiply has another constraint. 
Each ciphertext consists of $2$ or more polynomials.
\opmultiply of two ciphertexts with $k$ and $l$ polynomials 
yields a ciphertext with $k+l+1$ polynomials. 
Nevertheless, fewer the polynomials faster the \opmultiply, so
we enforce it to be the minimum:
\begin{equation}\label{eq:constraint-poly}
\begin{array}{c}
\forall{i}\ \evanpoly{\evaparam{n}{i}} = 2   \\
\text{ if } \opcode{n} \in \{\opmultiply\}
\end{array}
\end{equation}
\oprelinearize reduces the number of polynomials in a ciphertext to $2$. 
This constraint guarantees that 
any relinearization in the program would reduce the number of polynomials 
in a ciphertext from $3$ to $2$, 
thereby ensuring that only one public key is sufficient 
for all relinearizations in the program.
\oprelinearize can be inserted in the program in Figure~\ref{fig:example1}(d) 
to satisfy this constraint, 
as shown in Figure~\ref{fig:example1}(e).

Finally, 
we use $s_f$ to denote the maximum allowed rescale value in the 
rest of this paper
($\log_2 s_f$ is also the maximum bit size that can be used for encryption parameters), 
i.e.,
\begin{equation}\label{eq:constraint-rescale}
\begin{array}{c}
\constant{\evaparam{n}{2}} \leq s_f   \\
\text{ if } \opcode{n} \in \{\oprescale\}
\end{array}
\end{equation}
In the SEAL library, $s_f = 2^{60}$
(which enables a performant implementation 
by limiting scales to machine-sized integers).}


To summarize, FHE schemes (or libraries)
are tedious for a programmer to reason about, due to 
all their cryptographic constraints. 
Programmers find it even more tricky to satisfy the constraints
in a way that optimizes performance.
{\it The \eva compiler hides such cryptographic 
details from the programmer while optimizing the program}.

\subsection{Execution Flow of the Compiler}
\label{sec:eva-flow}

As mentioned in Section~\ref{sec:language}, the in-memory internal representation of the \eva compiler
is an {\bf Abstract Semantic Graph}, 
also known as a {\bf Term Graph}, of the input program. 
In the rest of this paper, we will use the term {\it graph} 
to denote an Abstract Semantic Graph. 
In this in-memory representation, each node can access both its parents 
and its children, 
and for each output, a distinct leaf node is added as a child.
It is straightforward to construct the graph from 
the \eva program and vice-versa, 
so we omit the details. 
We use the terms program and graph interchangeably 
in the rest of the paper. 

\begin{algorithm}[t]
\footnotesize
\caption{Execution of \eva compiler.}
\label{alg:eva-exec}
\SetKwInOut{Input}{Input}
\SetKwInOut{Output}{Output}
\Input{Program $P_i$ in \eva language}
\Input{Scales $S_i$ for inputs in $P_i$}
\Input{Desired scales $S_{d}$ for outputs in $P_i$}
\Output{Program $P_o$ in \eva language}
\Output{Vector $B_v$ of bit sizes}
\Output{Set $R_s$ of rotation steps}
$P_o =$ \textbf{\textsf{Transform}}($P_i$, $S_i$) \\
\If{\textbf{\textsf{Validate}}($P_o$) == Failed}{
    Throw an exception
}
$B_v =$ \textbf{\textsf{DetermineParameters}}($P_o$, $S_i$, $S_d$) \\
$R_s =$ \textbf{\textsf{DetermineRotationSteps}}($P_i$, $S_i$) 
\end{algorithm}

Algorithm~\ref{alg:eva-exec} 
presents the execution flow of the compiler. 
There are four main steps, namely 
transformation, validation, parameters selection, 
and rotations selection. 
The transformation step takes the input program 
and modifies it to satisfy the constraints of all instructions,
while optimizing it. 
In the next step, 
the transformed program is validated 
to ensure that no constraints are violated. 
If any constraints are violated, 
then the compiler throws an exception. 
By doing this, \eva ensures that executing the output program 
will never lead to a runtime exception thrown by the FHE library. 
Finally, for the validated output program, 
the compiler selects the bit sizes and the rotation steps 
that must be used to determine the encryption parameters 
and the public keys required for rotations respectively, 
before executing the output program. 
The transformation step involves rewriting the graph, 
which is described in detail in Section~\ref{sec:transformations}. 
The other steps only involve traversal of the graph 
(without changing it), which is described in 
Section~\ref{sec:analysis}.

\section{Transformations in \eva Compiler}
\label{sec:transformations}
\begin{figure*}[ht]
\centering
\small
\begin{minipage}{0.97\textwidth}
\begin{mathpar}
\inferrule*[LEFT= {\rm ALWAYS-RESCALE}]
{n \in \insts \\
\opcode{n} = \opmultiply \\
N_{ck}= \{ (n_c,k)\,|\,\evaparam{n_c}{k} = n \}}
{\insts \leftarrow \insts \cup \{n_s\} \\
\opcode{n_s} \leftarrow \oprescale \\
\evaparam{n_s}{1} \leftarrow n \\
\evaparam{n_s}{2} \leftarrow \min(\forall{j},\ \evascale{\evaparam{n}{j}}) \\
\forall{(n_c,k)\in N_{ck}},\ \evaparam{n_c}{k} \leftarrow n_s}

\inferrule*[LEFT= {\rm WATERLINE-RESCALE}]
{n \in \insts \\
\opcode{n} = \opmultiply \\
N_{ck}= \{ (n_c,k)\,|\,\evaparam{n_c}{k} = n \} \\
(\evascale{\evaparam{n}{1}} * \evascale{\evaparam{n}{2}}) / s_f \geq \max(\forall{n_j \in \{\consts, \inputs\}},\ \evascale{n_j})}
{\insts \leftarrow \insts \cup \{n_s\} \\
\opcode{n_s} \leftarrow \oprescale \\
\evaparam{n_s}{1} \leftarrow n \\
\evaparam{n_s}{2} \leftarrow s_f \\
\forall{(n_c,k)\in N_{ck}},\ \evaparam{n_c}{k} \leftarrow n_s}

\inferrule*[LEFT= {\rm LAZY-MODSWITCH}]
{n \in \insts \\
\opcode{n} \in \{\opadd, \opsub, \opmultiply\} \\
\evalevel{\evaparam{n}{i}} > \evalevel{\evaparam{n}{j}}}
{\insts \leftarrow \insts \cup \{n_m\} \\
\opcode{n_m} \leftarrow \opmodswitch \\
\evaparam{n_m}{1} \leftarrow \evaparam{n}{j} \\
\evaparam{n}{j} \leftarrow n_m}

\inferrule*[LEFT= {\rm EAGER-MODSWITCH}]
{n \in \{\insts, \consts, \inputs\} \\
\evaparam{n_c^1}{i} = n \\
\evaparam{n_c^2}{j} = n \\
\evarlevel{\evaparam{n_c^1}{i}} > \evarlevel{\evaparam{n_c^2}{j}} \\
N_{ck}= \{ (n_c,k)\,|\,\evaparam{n_c}{k} = n \land \evarlevel{\evaparam{n_c}{k}} = \evarlevel{\evaparam{n_c^2}{j}} \}}
{\insts \leftarrow \insts \cup \{n_m\} \\
\opcode{n_m} \leftarrow \opmodswitch \\
\evaparam{n_m}{1} \leftarrow n \\
\forall{(n_c,k)\in N_{ck}},\ \evaparam{n_c}{k} \leftarrow n_m}

\inferrule*[LEFT= {\rm MATCH-SCALE}]
{n \in \insts \\
\opcode{n} \in \{\opadd, \opsub\} \\
\evascale{\evaparam{n}{i}} > \evascale{\evaparam{n}{j}}}
{\insts \leftarrow \insts \cup \{n_t\} \\
\consts \leftarrow \consts \cup \{n_c\} \\
\constant{n_c} \leftarrow \evascale{\evaparam{n}{i}} / \evascale{\evaparam{n}{j}} \\
\opcode{n_t} \leftarrow \opmultiply \\
\evaparam{n_t}{1} \leftarrow \evaparam{n}{j} \\
\evaparam{n_t}{2} \leftarrow n_c \\
\evaparam{n}{j} \leftarrow n_t}

\inferrule*[LEFT= {\rm RELINEARIZE}]
{n \in \insts \\
\opcode{n} = \opmultiply \\
\evatype{\evaparam{n}{1}} = \evatype{\evaparam{n}{2}} = \typecipher \\
N_{ck}= \{ (n_c,k)\,|\,\evaparam{n_c}{k} = n \}}
{\insts \leftarrow \insts \cup \{n_l\} \\
\opcode{n_l} \leftarrow \oprelinearize \\
\evaparam{n_l}{1} \leftarrow n \\
\forall{(n_c,k)\in N_{ck}},\ \evaparam{n_c}{k} \leftarrow n_l}
\end{mathpar}
\end{minipage}
\caption{\revised{Graph rewriting rules (each rule is a transformation pass) in \eva ($s_f$: maximum allowed rescale value).}}
\label{fig:rewrite-rules}
\end{figure*}

In this section, 
we describe the key graph transformations 
in the \eva compiler. 
We first describe a general graph rewriting framework 
(Section~\ref{sec:graph-rewriting}). 
Then, we describe the graph transformation 
passes (Sections~\ref{sec:relinearize-insertion}
and~\ref{sec:rescale-insertion}).

\subsection{Graph Rewriting Framework}
\label{sec:graph-rewriting}

A graph transformation can be captured succinctly 
using graph {\it rewrite rules} 
(or term rewrite rules). 
These rules specify the conditional transformation of a subgraph 
(or an expression) 
and the graph transformation consists of 
transforming all applicable subgraphs (or expressions) 
in the graph (or program). 
The graph nodes have read-only properties 
like the opcode and number of parents.
In a graph transformation, 
some state or data may be stored on each node in the graph 
and the rewrite rules may read and update the state. 

The rewrite rules 
specify local operations on a graph 
and the graph transformation itself is composed 
of applying these operations wherever needed. 
The schedule in which these local operations are applied 
may impact the correctness or efficiency of the transformation. 
Consider two schedules: 
\begin{enumerate}
    \item Forward pass from roots to leaves of the graph:
    a node is scheduled for rewriting only 
    after all its parents have already been rewritten.
    \item Backward pass from leaves to roots of the graph:
    a node is scheduled for rewriting only 
    after all its children have already been rewritten.
\end{enumerate}
Note that the rewriting operation may not do any modifications 
if its condition does not hold.
In forward pass, state (or data) flows from parents to children. 
Similarly, in backward pass, state (or data) flows from 
children to parents.
In general, multiple forward or backward passes may be needed 
to apply the rewrite rules until quiescence (no change).

\eva includes a graph rewriting framework 
for arbitrary rewrite rules for a subgraph that
consists of a node along with its parents or children.
\revised{The rewrite rules for each graph transformation pass in \eva 
are defined in Figure~\ref{fig:rewrite-rules}. 
A single backward pass is sufficient for EAGER-MODSWITCH, 
while a single forward pass is sufficient for the rest. 
The rewrite rules assume the passes are applied in a specific order: 
WATERLINE-RESCALE, EAGER-MODSWITCH, MATCH-SCALE, and RELINEARIZE 
(ALWAYS-RESCALE and LAZY-MODSWITCH are not used but defined 
only for clarity). 
For the sake of exposition, 
we will first describe RELINEARIZE pass before describing the other passes.} 


\subsection{Relinearize Insertion Pass}
\label{sec:relinearize-insertion}

Each ciphertext is represented as 2 or more polynomials. 
Multiplying two ciphertexts each with 2 polynomials 
yields a ciphertext with 3 polynomials. 
The \oprelinearize instruction reduces a ciphertext 
to 2 polynomials. 
\revised{To satisfy Constraint~\ref{eq:constraint-poly},} 
\eva must insert \oprelinearize 
after \opmultiply of two nodes with \typecipher type 
and before another such \opmultiply.

The RELINEARIZE rewrite rule \revised{(Figure~\ref{fig:rewrite-rules})} 
is applied for a node $n$ 
only if it is a \opmultiply operation 
and if both its parents (or parameters) 
have \typecipher type.
The transformation in the rule 
inserts a \oprelinearize node $n_l$ 
between the node $n$ and its children. 
In other words, the new children of $n$ will be only $n_l$ 
and the children of $n_l$ will be the old children of $n$. 
\revised{For the example graph in Figure~\ref{fig:example1}(d), 
applying this rewrite rule transforms the graph into the one 
in Figure~\ref{fig:example1}(e).}

\revised{Optimal placement of relinearization is an 
NP-hard problem~\cite{HaoOptRelin}. 
Our relinearization insertion pass is a simple 
way to enforce Constraint~\ref{eq:constraint-poly}. 
More advanced relinearization insertion, 
with or without Constraint~\ref{eq:constraint-poly}, is left for future work.}


\subsection{Rescale and ModSwitch Insertion Passes}
\label{sec:rescale-insertion}

\paragraph{Goal:}
The \oprescale and \opmodswitch nodes (or instructions)
must be inserted such that they satisfy Constraint~\ref{eq:constraint-mod}, 
so the goal of the \oprescale and \opmodswitch insertion passes 
is to insert them such that the coefficient moduli of  
the parents of any \opadd and \opmultiply node are equal. 

\revised{While satisfying Constraint~\ref{eq:constraint-mod} is sufficient for correctness, 
performance depends on where \oprescale and \opmodswitch are inserted 
(as illustrated in Section~\ref{sec:constraints}).}
Different choices lead to different coefficient modulus 
$q = \{q_1, q_2, ..., q_r\}$, 
and consequently, different polynomial modulus $N$ 
for the roots (or inputs) to the graph (or program). 
Larger values of $N$ and $r$ increase the cost 
of every FHE operation and the memory of every ciphertext.
$N$ is a non-decreasing function of $Q = \prod_{i=1}^r q_i$ 
(i.e., if $Q$ grows, 
$N$ either remains the same or grows as well). 
Minimizing both $Q$ and $r$ is a hard problem to solve. 
However, reducing $Q$ is only impactful if it reduces $N$, 
which is unlikely
as the threshold of $Q$, for which $N$ increases, 
grows exponentially. 
Therefore, {\it the goal of \eva is to yield the optimal $r$}, 
which may or may not yield the optimal $N$.

\paragraph{Constrained-Optimization Problem:}
The only nodes that modify a ciphertext's coefficient modulus
are \oprescale and \opmodswitch nodes; 
that is, they are the only ones whose output ciphertext 
has a different coefficient modulus than that  
of their input ciphertext(s).
Therefore, the coefficient modulus of the output of a node 
depends only on the \oprescale and \opmodswitch nodes 
in the path from the root to that node.
To illustrate their relation, 
we define the term {\it rescale chain}.

\begin{definition}
Given a directed acyclic graph G = (V, E):

For $n_1, n_2 \in V$, $n_1$ is a \emph{parent} of $n_2$ 
if $\exists (n_1, n_2) \in E$.

A node $r \in V$ is a \emph{root} 
if $\evatype{r} = \typecipher$ and
$\nexists n \in V$ s.t. $n$ is a parent of $r$.
\end{definition}

\begin{definition}
Given a directed acyclic graph G = (V, E):

A \emph{path} $p$ from a node $n_0 \in V$ to a node $n \in V$ 
is a sequence of nodes 
$p_0, ..., p_l$ s.t. $p_0 = n_0$, $p_l = n$, and 
$\forall 0 \leq i < l, p_i \in V$ and $p_i$ is a parent of $p_{i+1}$. 
A {path} $p$ is said to be \emph{simple} 
if $\forall 0 < i < l, \opcode{p_i} \neq \oprescale$ 
and $\opcode{p_i} \neq \opmodswitch$. 
\end{definition}

\begin{definition}
Given a directed acyclic graph G = (V, E):

A \emph{rescale path} $p$ to a node $n \in V$ is a sequence of nodes 
$p_0, ..., p_l$ s.t. 
($\forall 0 \leq i \leq l, \opcode{p_i} \in \{\oprescale, \opmodswitch\}$),
$\exists$ a simple path from a root to $p_0$, 
$\exists$ a simple path from $p_l$ to $n$, 
($\forall 0 \leq i < l, \exists$ a simple path from $p_i$ to $p_{i+1}$), and 
($\opcode{n} = \oprescale$ or 
$\opcode{n} = \opmodswitch$) $\implies$ ($p_l = n$) . 

\label{def:rescale-chain}
A \emph{rescale chain} of a node $n \in V$ is a vector $c$ 
s.t. $\exists$ a rescale path $p$ to $n$ and 
($\forall 0 \leq i < |p|$,
($\opcode{p_i} = \opmodswitch \implies c_i = \infty$) and 
($\opcode{p_i} = \oprescale \implies c_i = \constant{\evaparam{p_i}{2}}$)).
Note that $\infty$ is used here to distinguish \opmodswitch from \oprescale in the rescale chain.

A rescale chain $c$ of a node $n$ and $c'$ of a node $n'$ are \emph{equal} 
if ($|c| = |c'|$ and 
($\forall 0 \leq i < |c|$,
$c_i = c_i'$ or $c_i = \infty$ 
or $c_i' = \infty$)).

A rescale chain $c$ of a node $n \in V$ is \emph{conforming} 
if $\forall$ rescale chain $c'$ of $n$, 
$c$ is equal to $c'$. 
\end{definition}

Note that all the roots in the graph have the same coefficient modulus. 
Therefore, for nodes $n_1$ and $n_2$, 
the coefficient modulus of the output of $n_1$ is equal to that of $n_2$
if and only if there exists conforming rescale chains for $n_1$ and $n_2$, and 
the conforming rescale chain of $n_1$ is equal to that of $n_2$. 
Thus, 
we need
to solve two problems simultaneously:
\begin{itemize}
\item Constraints: Ensure 
the conforming rescale chains of the parents of any \opmultiply or \opadd 
node are equal.
\item Optimization: Minimize the length of the rescale chain of every node.
\end{itemize}

\paragraph{Outline:} 
In general, the constraints problem can be solved in two steps:
\begin{itemize}
\item Insert \oprescale in a pass (to reduce exponential growth of scale and noise).
\item Insert \opmodswitch in another pass so that the constraints are satisfied.
\end{itemize}
The challenge is in solving this problem in this way, 
while yielding the desired optimization.

\paragraph{Always Rescale Insertion:}
A naive approach of inserting \oprescale is to insert it after 
every \opmultiply of \typecipher nodes. 
We call this approach as {\it always rescale} 
\revised{and define it in the ALWAY-RESCALE rewrite rule 
in Figure~\ref{fig:rewrite-rules}. 
Consider the example in Figure~\ref{fig:example1}(a). 
Applying this rewrite rule on this example yields the graph 
in Figure~\ref{fig:example1}(b). 
For some \opmultiply nodes (e.g., the bottom one), 
the conforming rescale chains of their parents do not exist or do not match. 
To satisfy these constraints, 
\opmodswitch nodes can be inserted appropriately, 
as shown in Figure~\ref{fig:example1}(c) 
(we omit defining this rule because it would require 
multiple forward passes). 
The conforming rescale chain length for the output 
is now more more  
than the multiplicative depth of the graph. 
Thus, {\it always rescale} and its corresponding modswitch insertion 
may lead to a larger coefficient modulus 
(both in the number of elements and their product)
than not inserting any of them. 
}


\paragraph{Insight:}
Consider that all the roots in the graph have the same scale $s$. 
\revised{For example in Figure~\ref{fig:example1}(a), 
let $\evascale{x} = 2^{30}$ instead of $2^{60}$.
Then,  
after {\it always rescale} 
(replace $2^{60}$ with $2^{30}$ in Figure~\ref{fig:example1}(b)),}
the only difference between the rescale chains of a node $n$ 
would be their length and not the values in it. 
\revised{This is the case even when roots may have different scales 
as long as all \oprescale nodes rescale by the same value $s$. 
Then,} a conforming rescale chain $c_n$ for $n$ can be obtained 
by adding \opmodswitch nodes in the smaller chain(s).
\revised{Thus, $|c_n|$
would not be greater than the multiplicative depth of $n$. 
The first key insight of \eva is that 
using the same rescale value for all \oprescale nodes 
ensures that 
$|c_o|$ cannot be greater than the multiplicative depth of $o$ 
(a tight upper bound).
The multiplicative depth of a node $n$ 
is not a tight lower bound for $|c_n|$, 
as shown in Figure~\ref{fig:example1}(d).
The second key insight of \eva is that 
using the maximum rescale value $s_f$ 
(satisfying Constraint~\ref{eq:constraint-rescale})
for all \oprescale nodes minimizes $|c_o|$ 
because it minimizes the number of \oprescale nodes in any path.}

\paragraph{Waterline Rescale Insertion:} 
Based on our insights, the value to rescale is fixed to 
$s_f$ ($= 2^{60}$ in SEAL).
That does not address the question of when to insert \oprescale nodes. 
\revised{If the scale after \oprescale becomes too small, 
then the computed message may lose accuracy irrevocably. 
We call the minimum required scale as {\it waterline}
and use $s_w$ to denote it. 
We choose $s_w$ to be maximum of the scales of all roots.} 
Consider a \opmultiply node $n$ whose scale after multiplication is $s_n$. 
Then, a \oprescale in inserted between $n$ and its children only if 
the scale after \oprescale is above the {\it waterline}, i.e., 
$(s_n / s_f) \geq s_w$.
We call this approach as {\it waterline rescale} 
\revised{and define the WATERLINE-RESCALE rewrite rule 
in Figure~\ref{fig:rewrite-rules}. 
This rule 
(assuming $s_w = 2^{30}$ instead of $2^{60}$)
transforms the graph in Figure~\ref{fig:example1}(a)
to the one in Figure~\ref{fig:example1}(d).}



\begin{figure}
\includegraphics[width=0.47\textwidth]{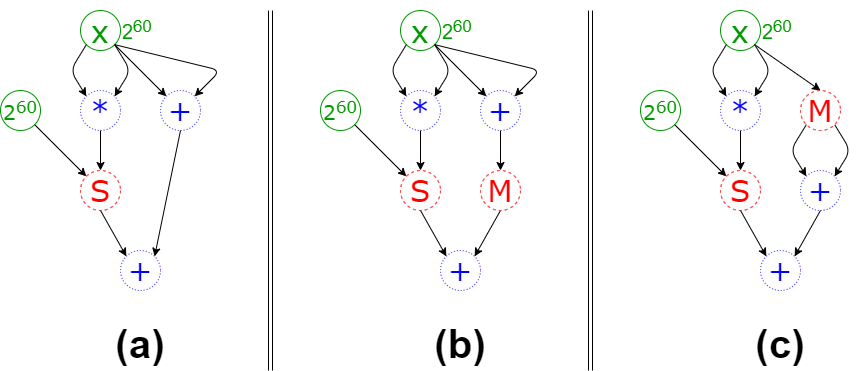}
\caption{\revised{$x^2+x+x$ in \evaend: (a) after WATERLINE-RESCALE (b) after WATERLINE-RESCALE \& LAZY-MODSWITCH; (c) after WATERLINE-RESCALE \& EAGER-MODSWITCH.}}
\label{fig:example3}
\end{figure}
    
\paragraph{ModSwitch Insertion:} 
\revised{For a node $n$,
let $\evalevel{n}$ denote its conforming rescale chain length.
Let $\evarlevel{n}$ denote the conforming rescale chain length of $n$
in the transpose graph.}
A naive way to insert \opmodswitch is to 
find a \opadd or \opmultiply node 
for which $level$ of the parents do not match  
and insert the appropriate number of \opmodswitch nodes 
between one of the parents and the node. 
We call this {\it lazy} insertion 
\revised{and define the LAZY-MODSWITCH rewrite rule 
in Figure~\ref{fig:rewrite-rules}}. 
We call inserting it at the earliest feasible edge in the graph
as {\it eager} insertion. 
\revised{The EAGER-MODSWITCH rewrite rule (Figure~\ref{fig:rewrite-rules}) 
finds a node for which $rlevel$ of the children do not match and 
inserts the appropriate number of \opmodswitch nodes between some of the children and itself. 
If the $rlevel$ of the roots do not match, 
then there is another rule 
(omitted in Figure~\ref{fig:rewrite-rules} for brevity) 
that inserts the appropriate \opmodswitch nodes 
between some of the roots and their children. 

Consider the $x^2+x+x$ example in Figure~\ref{fig:example3}(a). 
Applying the LAZY-MODSWITCH and EAGER-MODSWITCH rewrite rules yields
the graphs in Figures~\ref{fig:example3}(b) and (c) respectively. 
The operands of \opadd after eager insertion use a smaller coefficient modulus 
than after lazy insertion, 
so \opadd would be faster if eager insertion is used. 
Thus, eager insertion leads to similar or more efficient code 
than lazy insertion.}




\revised{
\paragraph{Matching Scales:}
As illustrated in Section~\ref{sec:constraints}, 
it is easy to match scales of parents of \opadd by 
multiplying one of the parents and $1$ with the appropriate scale. 
The MATCH-SCALE rewrite rule (Figure~\ref{fig:rewrite-rules}) 
takes this simple approach to satisfy Constraint~\ref{eq:constraint-scale} 
while avoiding introduction of any additional \oprescale or \opmodswitch. 
For the example graph in Figure~\ref{fig:example2}(a), 
applying this rewrite rule transforms the graph into the one 
in Figure~\ref{fig:example2}(c).

\paragraph{Optimality:} 
\eva selects encryption parameters 
(see Section~\ref{sec:analysis-passes}) 
s.t. $r = \max(\forall{o \in \{Outputs\}}, 1 + |c_o| + \lceil\frac{\evascale{o}*s_o}{s_f}\rceil$), 
where $s_o$ is the desired scale for the output $o$.
WATERLINE-RESCALE is the only pass that determines $|c_o|$ and $\evascale{o}$
for any output $o$ (neither LAZY-MODSWITCH nor MATCH-SCALE modify that). 
If $|c_o|$ is decreased by 1 (an element $s_f$ from $c_o$ is removed), 
then $\evascale{o}$ would increase by at least $s_f$, 
so it would not decrease $r$. 
Due to {\it waterline rescale}, $\evascale{o} < s_w * s_f$, 
so \oprescale cannot be inserted to reduce $\evascale{o}$ by at least $s_f$ 
(because the minimum required scale is $s_w$). 
Thus, \eva yields the minimal or optimal $r$.
}



\section{Analysis in \eva Compiler}
\label{sec:analysis}
In this section, 
we briefly describe our graph traversal framework 
(Section~\ref{sec:traversal}) 
and a few analysis passes 
(Section~\ref{sec:analysis-passes}). 

\subsection{Graph Traversal Framework and Executor}
\label{sec:traversal}

\evaend's graph traversal framework allows 
either a forward traversal or a backward traversal of the graph. 
In the forward traversal pass, 
a node is visited only after all its parents are visited. 
Similarly, in the backward traversal pass, 
a node is visited only after all its children are visited. 
Graph traversals do not modify the structure of the graph 
(unlike graph rewriting) 
but a state on each node can be maintained
 during the traversal. 
A single pass is sufficient to perform 
forward or backward data-flow analysis of the graph 
because the graph is acyclic. 
Execution of the graph 
is a forward traversal of the graph, 
so uses the same framework.

\paragraph{Parallel Implementation:} 
\revised{We implement an executor for the generated \eva program 
using the traversal framework.} 
A node is said to be {\it ready} or {\it active} 
if all its parents (or children) in forward (or backward) pass 
have already been visited. 
These active nodes can be scheduled to execute in parallel 
as each active node only updates its own state 
(i.e., there are no conflicts). 
\revised{For example in Figure~\ref{fig:example1}(e), 
the parents of the bottom \opmultiply can execute in parallel. 
Each FHE instruction (node) can take a significant amount of time to execute, 
so it is useful to exploit parallelism 
among FHE instructions.  
The \eva executor automatically parallelizes the generated \eva program}
by implementing a parallel graph traversal
using the Galois~\cite{galois,galoisproject} parallel library. 

\revised{A node is said to {\it retire} 
if all its children (or parents) in forward (or backward) pass 
have already been visited. 
The state for the retired nodes will no longer be accessed, 
so it can be reused for other nodes. 
In Figure~\ref{fig:example1}(e), 
the ciphertext for $x^2$ can be reused after the \oprelinearize is executed.
The \eva executor automatically reuses the memory used for encrypted messages, 
thereby reducing the memory consumed.}

\subsection{Analysis Passes}
\label{sec:analysis-passes}

\paragraph{Validation Passes:} 
We implement a validation pass for each of  
the constraints in Section~\ref{sec:constraints}. 
All are forward passes. 
\revised{The first pass computes the rescale chains for each node 
and asserts that it is conforming. 
It also asserts that the conforming rescale chains of parents 
of \opadd and \opmultiply match, 
satisfying Constraint~\ref{eq:constraint-mod}.
The second and third passes compute a $scale$ and $num\_polynomials$ 
for each node 
respectively
and assert that Constraint~\ref{eq:constraint-scale} 
and~\ref{eq:constraint-poly} is satisfied respectively.}
If any assertion fails, 
an exception in thrown at compile-time. 
Thus, these passes elide runtime exceptions thrown by SEAL.

\paragraph{Encryption Parameter Selection Pass:} 
Akin to encryption selection in CHET~\cite{CHET},
the encryption parameter selection pass in \eva 
computes the conforming {\it rescale chain} and the scale for each node. 
\revised{For each leaf or output $o$ after the pass, 
let $c_o$ be the conforming rescale chains of $o$ without $\infty$ in it 
and let $s'_o = s_o * \evascale{o}$, 
where $s_o$ is the desired output scale.
$s'_o$ is factorized into $s_0 * s_1 * ... * s_k$ 
such that $s_k$ is a power-of-two, $s_k \leq s_f$ ($= 2^{60}$ in SEAL), and 
$\forall 0 \leq i < k, s_i = s_f$. 
Let $|s'_o|$ denote the number of factors of $s'_o$. 
Then \eva finds the output $m$ with the maximum $|c_m| + |s'_m|$.
The factors of $s_m$ are appended to $c_m$
and $s_f$ (the {\it special prime}
that is consumed during encryption) is inserted 
at the beginning of $c_m$.}
For each element $s$ in $c_m$, 
$\log_2 s$ is applied to obtain 
a vector of bit sizes, which is then returned.

\paragraph{Rotation Keys Selection Pass:}
Similar to rotation keys selection in CHET~\cite{CHET}, 
\evaend's rotation keys selection pass  
\revised{computes and returns the set of unique step counts used 
among all \oprotateleft and \oprotateright nodes in the graph.}


\section{Frontends of \evaend}
\label{sec:apps}
The various transformations described so far for compiling an input EVA program into an executable EVA program make up the \emph{backend} in the EVA compiler framework. In this section, we describe two \emph{frontends} for EVA, that make it easy to write programs for EVA.

\begin{figure}
\begin{lstlisting}[language=python, backgroundcolor=\color{backgroundColour}]
from EVA import *
def sqrt(x):
  return x*constant(scale, 2.214) +
    (x**2)*constant(scale, -1.098) +
    (x**3)*constant(scale, 0.173)
program = Program(vec_size=64*64)
scale = 30
with program:
  image = inputEncrypted(scale)
  F = [[-1, 0, 1],
        [-2, 0, 2],
        [-1, 0, 1]]
  for i in range(3):
    for j in range(3):
      rot = image << (i*64+j)
      h = rot * constant(scale, F[i][j])
      v = rot * constant(scale, F[j][i])
      first = i == 0 and j == 0
      Ix = h if first else Ix + h
      Iy = v if first else Iy + v
  d = sqrt(Ix**2 + Iy**2)
  output(d, scale)
\end{lstlisting}
\caption{PyEVA program for Sobel filtering $64 \times 64$ images. The \lstinline{sqrt} function evaluates a 3rd degree polynomial approximation of square root.\label{fig:sobel}}
\end{figure}
  
\subsection{PyEVA}

We have built a general-purpose frontend for EVA as a DSL embedded into Python, called PyEVA. Consider the PyEVA program in Figure~\ref{fig:sobel} for Sobel filtering, which is a form of edge detection in image processing. 
The \lstinline{class Program} is a wrapper for the Protocol Buffer~\cite{protocolbuffer} format for EVA programs shown in Figure~\ref{fig:language}. 
It includes a context manager, such that inside a \lstinline{with program:} block all operations are recorded in \lstinline{program}. For example, the \lstinline{inputEncrypted} function inserts an input node of type \typecipher into the program currently in context and additionally returns an instance of \lstinline{class Expr}, which stores a reference to the input node. The expression additionally overrides Python operators to provide the simple syntax seen here. 


\subsection{EVA for Neural Network Inference}

CHET~\cite{CHET} is a compiler for evaluating neural networks on encrypted inputs. The CHET compiler receives a neural network as a graph of high-level tensor operations, and through its kernel implementations, analyzes and executes these neural networks against FHE libraries. 
CHET lacks a proper backend and operates more as an interpreter coupled with automatically chosen high-level execution strategies.

We have obtained the CHET source code and modified it to use the EVA compiler as a backend. CHET uses an interface called \emph{Homomorphic Instruction Set Architecture} (HISA) as a common abstraction for different FHE libraries. In order to make CHET generate EVA programs, we introduce a new HISA implementation that instead of calling homomorphic operations inserts instructions into an EVA program. This decouples the generation of the program from its execution. We make use of CHET's data layout selection optimization, but not its encryption parameter selection functionality, as this is already provided in EVA. Thus, EVA subsumes CHET.


\section{Experimental Evaluation}
\label{sec:eval}
\newcommand{\mnistall}{LeNet-5-like\xspace}
\newcommand{\mnistsmall}{LeNet-5-small\xspace}
\newcommand{\mnistmed}{LeNet-5-medium\xspace}
\newcommand{\mnistlarge}{LeNet-5-large\xspace}
\newcommand{\mnistlargehf}{LeNet-5-large-hf\xspace}
\newcommand{\mnetname}{Industrial\xspace}
\newcommand{\squeezenet}{SqueezeNet-CIFAR\xspace}

In this section, 
we first describe our experimental setup 
(Section~\ref{sec:exp-setup}). 
We then describe our evaluation of 
homomorphic neural network inference 
(Section~\ref{sec:eval-dnn}) 
and homomorphic arithmetic, 
statistical machine learning, and
image processing applications 
(Section~\ref{sec:eval-other}).

\subsection{Experimental Setup}
\label{sec:exp-setup}

All experiments were 
conducted on a 4 socket machine with Intel Xeon Gold 5120
2.2GHz CPU with 56 cores (14 cores per socket) and 190GB memory.
Our evaluation of all applications 
uses GCC 8.1 and SEAL v3.3.1~\cite{sealcrypto}, 
which implements the 
RNS variant of the CKKS scheme~\cite{SAC:CHKKS18}.
All experiments use the default 128-bit security level. 

We evaluate a simple arithmetic application 
to compute the path length in 3-dimensional space. 
We also evaluate
applications in statistical machine learning, 
image processing, and deep neural network (DNN) inferencing 
using the frontends that we built 
on top of \eva (Section~\ref{sec:apps}). 
For DNN inferencing, we compare \eva with the 
state-of-the-art compiler for homomorphic DNN inferencing, 
CHET~\cite{CHET}, 
which has been shown to outperform hand-tuned codes. 
For the other applications, no suitable compiler exists 
for comparison. 
Hand-written codes also do no exist 
as it is very tedious to write them manually.
We evaluate these applications using \eva to show that 
\eva yields good performance with little programming effort.
\revised{For DNN inferencing, 
the accuracy reported is for all test inputs, 
whereas all the other results reported are an average 
over the first 20 test inputs. 
For the other applications, 
all results reported are an average over 20 different 
randomly generated inputs}.

\begin{table}[t]
\centering
\small
\caption{Deep Neural Networks used in our evaluation.}
\label{fig:networks}
\begin{tabularx}{\columnwidth}{@{}Xrrrrr@{}}
\toprule
\multirow{2}{*}{\bf Network} & \multicolumn{3}{c}{\bf No. of layers} & \multicolumn{1}{c}{\bf \# FP} & \multicolumn{1}{c@{}}{\bf Accu-}\\
\cmidrule{2-4}
    & {\bf Conv} & {\bf FC} & {\bf Act} & {\bf operations} & \multicolumn{1}{c@{}}{\bf racy(\%)} \\
\midrule
LeNet-5-small & 2 & 2 & 4 & 159960 & 98.45 \\
LeNet-5-medium & 2 & 2 & 4 & 5791168 & 99.11 \\
LeNet-5-large & 2 & 2 & 4 & 21385674 & 99.30 \\
Industrial & 5 & 2 & 6 & - & - \\
SqueezeNet-CIFAR & 10 & 0 & 9 & 37759754 & 79.38 \\
\bottomrule
\end{tabularx}
\end{table}

\begin{table}[t]
\centering
\small
\caption{Programmer-specified input and output scaling factors used for both CHET and EVA, and the accuracy of homomorphic inference in CHET and EVA (all test inputs).}
\label{fig:scales}
\begin{tabularx}{\columnwidth}{@{}X@{}@{}c@{}c@{}c@{ }c@{ }c@{}c@{}}
\toprule
\multicolumn{1}{@{}l@{}}{\multirow{2}{*}{\textbf{Model}}} & \multicolumn{3}{@{}c@{}}{\textbf{Input Scale ($\log P$)}}                                                                                                      & \multicolumn{1}{@{ }c@{ }}{\textbf{Output}} & \multicolumn{2}{@{ }c@{}}{{\textbf{\revised{Accuracy(\%)}}}} \\
\cmidrule{2-4}
\cmidrule{6-7}
                                & \multicolumn{1}{@{}c@{}}{\textbf{Cipher}} & \multicolumn{1}{@{ }c@{}}{\textbf{Vector}} & \multicolumn{1}{@{ }c@{}}{\textbf{Scalar}} & \multicolumn{1}{@{ }c@{ }}{\textbf{Scale}}  & \multicolumn{1}{@{ }c@{}}{{\textbf{\revised{CHET}}}} & \multicolumn{1}{@{ }c@{}}{{\textbf{EVA}}}                                       \\
\midrule
LeNet-5-small                                       & 25                                  & 15                                 & 10                                    & 30                                  & \revised{98.42} & 98.45                                                      \\
LeNet-5-medium                                      & 25                                  & 15                                 & 10                                    & 30                                  & \revised{99.07} & 99.09                                                     \\
LeNet-5-large                                       & 25                                  & 20                                 & 10                                    & 25                                  & \revised{99.34} & 99.32                                                            \\
Industrial                                              & 30                                  & 15                                 & 10                                    & 30                              & \revised{-} & -                                                            \\
SqueezeNet-CIFAR                                    & 25                                  & 15                                 & 10                                    & 30                                  & \revised{79.31} & 79.34 \\
\bottomrule                                                          
\end{tabularx}
\end{table}

\subsection{Deep Neural Network (DNN) Inference}
\label{sec:eval-dnn} 
    
\paragraph{Networks:}
We evaluate five deep neural network (DNN) architectures for image
classification tasks that are summarized in Table~\ref{fig:networks}:
\begin{itemize}
\item
The three {\bf LeNet-5} networks are all 
for the MNIST~\cite{mnist}
dataset, which vary in the number of neurons. 
The largest one matches the
one used in the TensorFlow's tutorials~\cite{lenet5like}. 
\item
{\bf \mnetname} is a 
network from an industry partner
for privacy-sensitive binary classification of images. 
\item
{\bf \squeezenet} is a 
network for the CIFAR-10 dataset~\cite{cifar10}
that uses 4 Fire-modules~\cite{squeezenet-cifar10} 
and 
follows the SqueezeNet~\cite{SqueezeNet} architecture. 
\end{itemize} 

We obtain these networks (and the models) 
from the authors of CHET, 
so they match the networks evaluated in their paper~\cite{CHET}.
\mnetname is a FHE-compatible neural network 
that is proprietary, 
so the authors gave us only the network structure 
without the trained model (weights) or the test datasets. 
We evaluate this network using randomly generated 
numbers (between -1 and 1) for the model and the images.
All the other networks were made FHE-compatible by CHET authors
using average-pooling and polynomial activations 
instead of max-pooling and ReLU activations.
Table~\ref{fig:networks} lists the accuracies 
we observed for these networks using 
unencrypted inference on the test datasets.
We evaluate encrypted image inference with a batch size of 1 
(latency).

\paragraph{Scaling Factors:} 
The scaling factors, or scales in short, 
must be chosen by the user. 
For each network (and model), 
we use CHET's profiling-guided optimization 
on the first 20 test images 
to choose the input scales as well as the 
desired output scale. 
There is only one output but there are many inputs.
For the inputs, we choose one scale each for 
\typecipher, \typevector, and \typescalar 
inputs. 
Both CHET and \eva use the same scales, 
as shown in Table~\ref{fig:scales}. 
The scales impact both performance and accuracy. 
\revised{We evaluate CHET and \eva on all test images using these scales 
and report the accuracy achieved by fully-homomorphic inference 
in Table~\ref{fig:scales}.
There is negligible difference between their accuracy 
and the accuracy of unencrypted inference 
(Table~\ref{fig:networks}).} 
Higher values of scaling factors may improve the accuracy, 
but will also increase the latency of homomorphic inference.

\begin{table}[t]
\centering
\small
\caption{Average latency (s) of CHET and EVA on 56 threads.}
\label{fig:latency}
\begin{tabularx}{\columnwidth}{@{}Xrrr@{}}
\toprule
\textbf{Model} & \textbf{CHET} & \textbf{EVA} & \textbf{Speedup from EVA} \\
\midrule
LeNet-5-small                      & 3.7                               & 0.6                              & 6.2                                  \\
LeNet-5-medium                     & 5.8                               & 1.2                              & 4.8                                  \\
LeNet-5-large                      & 23.3                              & 5.6                              & 4.2                                  \\
Median                             & 70.4                              & 9.6                              & 7.3                                  \\
SqueezeNet-CIFAR                   & 344.7                             & 72.7                             & 4.7             \\
\bottomrule                    
\end{tabularx}
\end{table}

\begin{figure}
\includegraphics[width=0.47\textwidth]{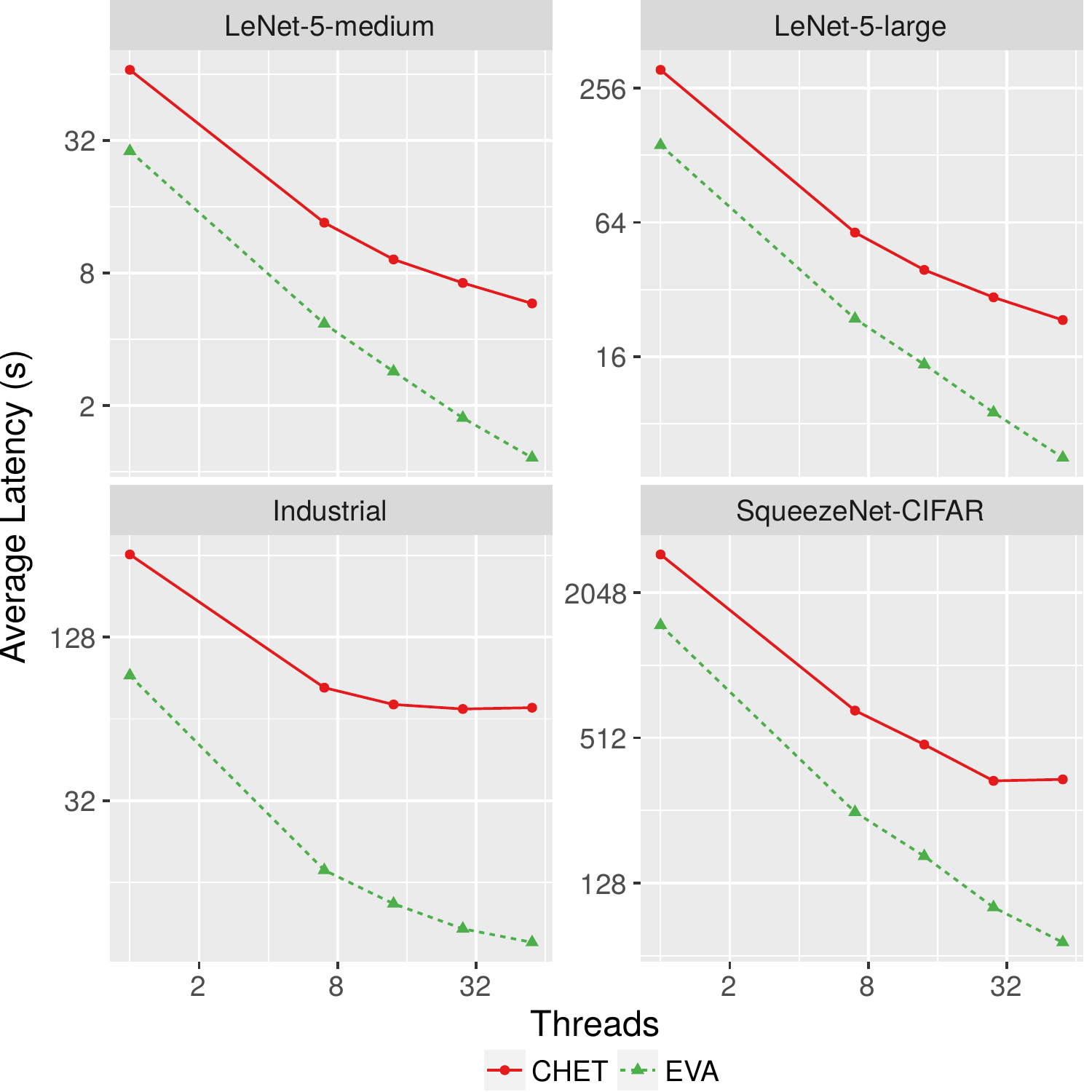}
\caption{Strong scaling of CHET and \eva (log-log scale).}
\label{fig:scaling}
\end{figure}

\paragraph{Comparison with CHET Compiler:} 
Table~\ref{fig:latency} shows that \eva 
is at least $4\times$ faster than CHET on 56 threads for all networks. 
Note that the average latency of CHET is slower than that 
reported in their paper~\cite{CHET}. 
This could be due to differences in the experimental setup. 
The input and output scales they use are different, 
so is the SEAL version (3.1 vs. 3.3.1). 
We suspect the machine differences to be the primary reason 
for the slowdown 
because they use smaller number of heavier cores 
(16 3.2GHz cores vs. 56 2.2GHz cores). 
In any case, our comparison of CHET and \eva is fair 
because both use the same input and output scales, SEAL version, 
Channel-Height-Width (CHW) data layout, and hardware. 
\revised{Both CHET and EVA perform similar 
encryption parameters and rotation keys selection.
The differences between CHET and \eva are solely 
due to the benefits that accrue from \evaend's low-level optimizations. 

CHET relies on an expert-optimized library of homomorphic 
tensor kernels, 
where each kernel (1) includes FHE-specific instructions 
and (2) is explicitly parallelized. 
However, even experts cannot optimize or parallelize across different kernels 
as that information is not available to them. 
In contrast, 
\eva uses a library of vectorized tensor kernels and automatically
(1) inserts FHE-specific instructions using global analysis and 
(2) parallelizes the execution of different instructions across kernels.
Due to these optimizations, 
{\it \eva is on average $5.3\times$ faster than CHET}. 
On a single thread (Figure~\ref{fig:scaling}), 
\eva is on average $2.3\times$ faster than CHET and 
this is solely due to better placement of FHE-specific instructions. 
The rest of the improvement on 56 threads ($2.3\times$ on average) is 
due to better parallelization in \evaend.

Both CHET and \eva have similar \oprelinearize placement. 
However, they differ in the placement of the other FHE-specific instructions ---
\oprescale and \opmodswitch. 
These instructions directly impact the encryption parameters 
(both CHET and \eva use a similar encryption parameter selection pass).}
We report the encryption parameters 
selected by CHET and \eva in Table~\ref{fig:params}. 
\eva selects much smaller coefficient modulus, 
both in terms of the number of elements \revised{$r$} in it 
and their product \revised{$Q$}. Consequently, the polynomial modulus \revised{$N$}
is one power-of-2 lower in all networks, except \mnistlarge. 
\revised{Reducing $N$ and $r$} reduces the cost (and the memory) 
of each homomorphic operation (and ciphertext) 
significantly.
In CHET, \oprescale and \opmodswitch used by 
the experts for a given tensor kernel may be sub-optimal 
for the program. 
On the other hand, \eva performs global (inter-procedural) 
analysis to minimize the length of the coefficient modulus, 
yielding much smaller encryption parameters.

\revised{To understand the differences in parallelization,
we evaluated CHET and \eva} on 1, 7, 14, 28, and 56 threads. 
Figure~\ref{fig:scaling} shows the strong scaling. 
We omit \mnistsmall because it takes too little time, 
even on 1 thread.
It is apparent that \eva scales much better than CHET. 
The parallelization in CHET is within a tensor operation 
or kernel using OpenMP. 
Such static, {\it bulk-synchronous} schedule limits 
the available parallelism. 
In contrast, 
\eva dynamically schedules the directed acyclic graph 
of \eva (or SEAL) operations asynchronously. 
Thus, it exploits the parallelism available 
across tensor kernels, resulting in much better scaling. 
The average speedup of \eva on 56 threads over \eva on 1 thread 
is $18.6\times$ (excluding \mnistsmall).

\begin{table}[t]
\centering
\small
\caption{Encryption parameters selected by CHET and EVA (where $Q = \prod_{i=1}^r Q_i$).}
\label{fig:params}
\begin{tabularx}{\columnwidth}{@{}X@{ }r@{\quad}r@{\quad}rr@{\quad}r@{\quad}r}
\toprule
\multicolumn{1}{@{}l}{\multirow{2}{*}{\textbf{Model}}} & \multicolumn{3}{c}{\textbf{CHET}}                             & \multicolumn{3}{c}{\textbf{EVA}}                                \\
\cmidrule(lr){2-4}\cmidrule(lr){5-7}
                                & \textbf{$\log_2 N$} & \textbf{$\log_2 Q$} & \textbf{$r$} & \textbf{$\log_2 N$} & \textbf{$\log_2 Q$} & \textbf{$r$} \\
\midrule
LeNet-5-small                   & 15              & 480                   & 8                   & 14              & 360                   & 6                     \\
LeNet-5-medium                  & 15              & 480                   & 8                   & 14              & 360                   & 6                     \\
LeNet-5-large                   & 15              & 740                   & 13                  & 15              & 480                   & 8                     \\
Industrial                          & 16              & 1222                  & 21                  & 15              & 810                   & 14                    \\
SqueezeNet-CIFAR                & 16              & 1740                  & 29                  & 16              & 1225                  & 21       \\            
\bottomrule
\end{tabularx}
\end{table}

\paragraph{Comparison with Hand-Written LoLa:} 
LoLa~\cite{BrutzkusGE19} 
implements hand-tuned homomorphic inference for 
neural networks, but the networks they implement 
are different than the ones we evaluated (and the ones in CHET). 
Nonetheless, they implement networks for the MNIST 
and CIFAR-10 datasets. 

For the MNIST dataset, 
LoLa implements the highly-tuned CryptoNets~\cite{cryptonets} network 
(which is similar in size to \mnistsmall). 
This implementation has an average latency of $2.2$ seconds 
and has an accuracy of $98.95\%$. 
\eva takes only $1.2$ seconds on a much larger network, \mnistmed, 
with a better accuracy of $99.09\%$.
For the CIFAR-10 dataset, 
LoLa implements a custom network which takes $730$ seconds 
and has an accuracy of $74.1\%$. 
\eva takes only $72.7$ seconds on a much larger network 
with a better accuracy of $79.34\%$. 

LoLa uses SEAL 2.3 
(which implements BFV~\cite{FV12}) 
which is less efficient than SEAL 3.3.1 
(which implements RNS-CKKS~\cite{SAC:CHKKS18})
but much more easier to use.
\eva is faster because 
it exploits a more efficient FHE 
scheme which is much more difficult to manually write code for.
Thus, {\it \eva outperforms even highly tuned expert-written 
implementations like LoLa with very little programming effort}. 

\begin{table}[t]
\centering
\small
\caption{Compilation, encryption context (context), encryption, and decryption time for EVA.}
\label{fig:compile-time}
\begin{tabularx}{\columnwidth}{@{}X@{ }@{}r@{~}r@{ }r@{ }r@{}}
\toprule
\multicolumn{1}{@{}l@{}}{\multirow{2}{*}{\textbf{Model}}} & \multicolumn{4}{c}{\textbf{Time (s)}}                                                                                                                                                                                      \\
\cmidrule{2-5}
\multicolumn{1}{c@{}}{}                                
& \multicolumn{1}{@{ }l@{}}{\textbf{Compilation}} & \multicolumn{1}{@{ }l@{}}{\textbf{Context}} & \multicolumn{1}{@{ }l@{}}{\textbf{Encrypt}} & \multicolumn{1}{@{ }l@{}}{\textbf{Decrypt}} \\
\midrule
LeNet-5-small                                       
                                            & 0.14                                     & 1.21                                            & 0.03                                 & 0.01                                 \\
LeNet-5-medium                                      
                                            & 0.50                                     & 1.26                                            & 0.03                                 & 0.01                                 \\
LeNet-5-large                                       
                                            & 1.13                                     & 7.24                                            & 0.08                                 & 0.02                                 \\
Industrial                                              
                                            & 0.59                                     & 15.70                                           & 0.12                                 & 0.03                                 \\
SqueezeNet-CIFAR                                    
                                            & 4.06                                     & 160.82                                          & 0.42                                 & 0.26  \\
\bottomrule                              
\end{tabularx}
\end{table}

\begin{table}[t]
\centering
\small
\caption{Evaluation of \eva for fully-homomorphic arithmetic, 
statistical machine learning, and image processing applications on 1 thread (LoC: lines of code).}
\label{tbl:otherapps}
\begin{tabularx}{\columnwidth}{@{}Xrrr@{}}
\toprule
\textbf{Application} & \textbf{Vector Size} & \textbf{LoC} & \textbf{Time (s)} \\
\midrule
3-dimensional Path Length                           & 4096                                     & 45                                         & 0.394                                                   \\
Linear Regression                        & 2048                                     & 10                                         & 0.027                                                   \\
Polynomial Regression                    & 4096                                     & 15                                         & 0.104                                                   \\
Multivariate Regression                  & 2048                                     & 15                                         & 0.094                                                   \\
Sobel Filter Detection                   & 4096                                     & 35                                         & 0.511                                                   \\
Harris Corner Detection                  & 4096                                     & 40                                         & 1.004          \\
\bottomrule                                        
\end{tabularx}
\end{table}

\paragraph{Compilation Time:} 
We present the compilation time, encryption context time, 
encryption time, and decryption time for all networks 
in Table~\ref{fig:compile-time}. 
The encryption context time includes the time 
to generate the public key, the secret key, 
the rotation keys, and the relinearization keys. 
This can take a lot of time, especially for large $N$, 
like in \squeezenet. 
Compilation time, encryption time, and decryption time 
are negligible for all networks.

\subsection{Arithmetic, Statistical Machine Learning, 
and Image Processing}
\label{sec:eval-other}

We implemented several applications using Py\eva. 
To illustrate a simple arithmetic application, 
we implemented an application 
that computes the length of a given 
encrypted 3-dimensional path. 
This computation can be used as a kernel 
in several applications like in
secure fitness tracking on mobiles. 
For statistical machine learning, 
we implemented linear regression, 
polynomial regression, and multi-variate regression 
on encrypted vectors. 
For image processing, 
we implemented Sobel filter detection and 
Harris corner detection on encrypted images.
All these implementations took very few lines of code
($< 50$), as shown in Table~\ref{tbl:otherapps}.

Table~\ref{tbl:otherapps} shows the execution time 
of these applications on encrypted data using 1 thread. 
Sobel filter detection takes half a second and 
Harris corner detection takes only a second. 
The rest take negligible time. 
We believe \revised{\it Harris corner detection is one of the most 
complex programs that have been evaluated using CKKS}. 
\eva enables writing advanced applications in various domains 
with little programming effort, 
while providing excellent performance.

\section{Related Work}
\label{sec:related}

\revised{
\paragraph{Libraries for FHE} 
SEAL~\cite{sealcrypto} implements RNS variants of two FHE schemes: 
BFV~\cite{FV12} and CKKS~\cite{AC:CKKS17,SAC:CHKKS18}. 
HElib~\cite{helib} implements two FHE schemes: 
BGV~\cite{ITCS:BraGenVai12} and CKKS. 
PALISADE~\cite{palisade} is a framework that provides a general API 
for multiple FHE schemes including BFV, BGV, and CKKS. 
For BFV and CKKS, PALISADE is similar to SEAL as it only implements lower-level 
FHE primitives. 
On the other hand, \eva language abstracts batching-compatible FHE schemes like 
BFV, BGV, and CKKS 
while hiding cryptographic details from the programmer.
Although \eva compiler currently generates code 
targeting only CKKS implementation in SEAL, 
it can be adapted to target other 
batching-compatible FHE scheme implementations or FHE libraries.

\paragraph{General-Purpose Compilers for FHE}
To reduce the burden of writing FHE programs, 
general-purpose compilers have been proposed that target different FHE libraries.  
These compilers share many of the same goals as ours.
Some of these compilers support general-purpose languages like Julia (cf.~\cite{ArcherTDMPRR19}), C++ (cf.~\cite{cingulata}), and R (cf.~\cite{aslett2015review}), whereas 
ALCHEMY~\cite{alchemy} is the only one that 
provides its own general-purpose language. 
Unlike \evaend, 
none of these languages are amenable to be a target for 
domain-specific compilers like CHET~\cite{CHET} 
because these languages do not support rotations on 
fixed power-of-two sized vectors.
Nonetheless, techniques in these compilers 
(such as ALCHEMY's static type safety and error rate analysis)
are orthogonal to 
our contributions in this paper and can be incorporated in EVA.

All prior general-purpose compilers target 
(libraries implementing) either the BFV scheme~\cite{FV12} or the BGV scheme~\cite{ITCS:BraGenVai12}.
In contrast, \eva targets 
(libraries implementing) the recent CKKS scheme~\cite{AC:CKKS17,SAC:CHKKS18},  
which is much more difficult to write or generate code for 
(compared to BFV or BGV). 
For example, 
ALCHEMY supports the BGV scheme and 
would require significant changes to capture the semantics 
(e.g., \oprescale) of CKKS.
ALCHEMY always inserts \opmodswitch 
after every ciphertext-ciphertext multiplication (using local analysis), 
which is not optimal for BGV (or BFV) and would not be correct for CKKS.
\eva is the first general-purpose compiler for CKKS and it 
uses a graph rewriting framework to insert \oprescale 
and \opmodswitch operations correctly (using global analysis)
so that the modulus chain length is optimal. 
These compiler passes in \eva 
can be incorporated in other general-purpose compilers (to target CKKS). 
}

\paragraph{\revised{Domain-Specific Compilers for FHE}}
\revised{Some prior compilers 
for DNN inferencing~\cite{CHET,BoemerLCW19,BoemerCCW19} target CKKS.  
CHET~\cite{CHET} is a compiler for tensor programs that 
automates the selection of {\it data layouts} for mapping tensors to 
vectors of vectors.} 
The nGraph-HE~\cite{BoemerLCW19} project introduced an extension to the Intel nGraph~\cite{ngraph} deep learning compiler that allowed data scientists to make use of FHE with minimal code changes. 
The nGraph-HE compiler uses run-time optimization (e.g., detection of special plaintext values) and compile-time optimizations (e.g., use of ISA-level parallelism, graph-level optimizations).
nGraph-HE2~\cite{BoemerCCW19} is an extension of nGraph-HE 
that uses a hybrid computational model -- the server interacts with the client to perform non-HE compatible operations, which increases the communication overhead.
Moreover, unlike CHET and \evaend, neither nGraph-HE nor nGraph-HE2 automatically select encryption parameters.

\revised{
To hide the complexities of FHE operations, 
all existing domain-specific compilers 
rely on a runtime of high-level 
kernels which can be optimized by experts. 
However, experts are limited to information within a single kernel 
(like convolution) 
to optimize insertion of FHE-specific operations 
and to parallelize execution. 
In contrast, \eva  
optimizes insertion of FHE-specific operations by using global analysis
and parallelizes FHE operations across kernels transparently.
Therefore, CHET, nGraph-HE, and nGraph-HE2 can target \eva 
instead of the FHE scheme directly to benefit 
from such optimizations and  
we demonstrated this for CHET.}

\paragraph{Compilers for MPC}
Multi-party computation (MPC)~\cite{STOC:GolMicWig87,FOCS:Yao86} is another technique for privacy-preserving computation. 
The existing MPC compilers~\cite{SP:HHNZ19} are mostly general-purpose and even though it is possible to use them for deep learning applications, it is hard to program against a general-purpose interface.
The EzPC compiler~\cite{ChandranGRST19} is a machine learning compiler that combines arithmetic sharing and garbled circuits and operates in a two-party setting. 
EzPC uses ABY~\cite{CCS:MohRin18} as a cryptographic backend. 

\paragraph{Privacy-Preserving Deep Learning}
CryptoNets~\cite{cryptonets}, one of the first systems for neural network inference using FHE and the consequent work on LoLa~\cite{BrutzkusGE19}, a low-latency CryptoNets, show the ever more practical use of FHE for deep learning. 
CryptoNets and LoLa however use kernels for neural networks that directly translate the operations to the cryptographic primitives of the FHE schemes.
There are also other algorithms and cryptosystems specifically for deep learning that rely on FHE (CryptoDL~\cite{crypto-dl}, Chabanne et~al.~\cite{Chabanne2017}, Jiang et~al.~\cite{CCS:JKLS18}), MPC (Chameleon~\cite{ASIACCS:RWTSSK18}, DeepSecure~\cite{Rouhani:2018:DSP:3195970.3196023}, SecureML~\cite{SP:MohZha17}), oblivious protocols (MiniONN~\cite{CCS:LJLA17}), or on hybrid approaches (Gazelle~\cite{USENIX:JuvVaiCha18}, SecureNN~\cite{PoPETS:WagGupCha19}).
None of these provide the flexibility and the optimizations of a compiler approach.

\section{Conclusions}
\label{sec:conclusions}
This paper introduces a new language and 
intermediate representation called 
Encrypted Vector Arithmetic (\eva)
for general-purpose Fully-Homomorphic Encryption (FHE) computation. 
\eva includes a Python frontend that can be used to write 
advanced programs with little programming effort, and it 
hides all the cryptographic details from the programmer. 
\eva includes an optimizing compiler that generates 
correct, secure, and efficient code, targeting the
state-of-the-art SEAL library.
EVA is also designed for easy targeting of domain specific languages.
The state-of-the-art neural network inference compiler CHET, when re-targeted 
onto EVA, outperforms its unmodified version by $5.3\times$ on average. 
\eva provides a solid foundation for a richer variety of FHE applications 
as well as domain-specific compilers and 
\revised{auto-vectorizing compilers for 
computing on encrypted data}.

\section*{Acknowledgments}
This  research  was  supported  by  the  NSF  grants  1406355, 1618425, 1705092, 1725322,  
and  by  the  DARPA  contracts  FA8750-16-2-0004 and FA8650-15-C-7563. 
We thank Keshav Pingali for his support.
We thank the anonymous reviewers and in particular our
shepherd, Petar Tsankov, for their many suggestions in improving
this paper.

\balance
\bibliography{abbrev0,crypto,references,chet_references,iss}


\end{document}